\documentclass[fleqn,usenatbib]{mnras}
\usepackage[T1]{fontenc}
\DeclareRobustCommand{\VAN}[3]{#2}
\let\VANthebibliography\thebibliography
\def\thebibliography{\DeclareRobustCommand{\VAN}[3]{##3}\VANthebibliography}
\usepackage{bm}
\usepackage{graphicx}
\usepackage{amsmath}
\usepackage{amssymb}
\newcommand{\be}{\begin{equation}}
\newcommand{\ee}{\end{equation} }
\newcommand{\ba}{\begin{eqnarray}}
\newcommand{\ea}{\end{eqnarray}}

\newcommand{\bnabla}{\mbox{\boldmath$\nabla$}}

\newcommand{\bB}{{\bm B}}
\newcommand{\wB}{\widetilde{B}}
\newcommand{\wbB}{\widetilde{{\bm B}}}
\newcommand{\bE}{{\bm E}}
\newcommand{\wE}{\widetilde{E}}
\newcommand{\wbE}{\widetilde{{\bm E}}}
\newcommand{\bk}{{\bm k}}
\newcommand{\wk}{\widetilde{k}}
\newcommand{\wbk}{\widetilde{{\bm k}}}
\newcommand{\wom}{\widetilde{\omega}}
\newcommand{\bx}{{\bm x}}
\newcommand{\wbeta}{\widetilde{\beta}}
\newcommand{\womega}{\widetilde{\omega}}

\title[Direct Emission of Strong Radio Pulses]{Direct Emission of Strong Radio Pulses during \\ Magnetar Flares}
\author[C. Thompson]{Christopher Thompson$^{1}$\thanks{E-mail:thompson@cita.utoronto.ca}
\\
$^{1}$ Canadian Institute for Theoretical Astrophysics, 60 St. George St., Toronto, ON M5S 3H8, Canada.}		

\date{MNRAS in press}

\pubyear{2022}
\begin{document}
\label{firstpage}
\pagerange{\pageref{firstpage}--\pageref{lastpage}}
\maketitle

\begin{abstract}
  The emission of intense radio pulses by flaring magnetars is investigated.
  Small-scale current gradients can be imprinted into a strongly magnetized outflow by the same
  processes that source fireball radiation in the closed magnetosphere.
  This structure arises from a combination of crustal yielding, internal tearing, and turbulent cascade.
  We consider the quasi-linear development of weak, small-scale currents
  as (i) they are stretched out and frozen by relativistic expansion and then (ii) 
  pass through a shock.  In particular, we derive the amplitudes of the ordinary and
  fast waves that emerge downstream of a relativistically magnetized shock in response to a
  mode that is frozen into the upstream flow (a frozen Alfv\'en wave or entropy wave).
  An upstream mode with comoving wavelength exceeding the skin depth can linearly
  convert to a secondary mode propagating above the plasma frequency.
  A simple and accurate treatment of shocks with extreme magnetization is developed, and the formation
  of internal shocks in the outflow from a bursting, rotating magnetar is outlined.  The emission process
  described here does not require a strong shock or cool $e^\pm$ pairs (in contrast with the
  electromagnetic maser shock instability).  In some cases, a high-frequency wave is reflected
  back to the observer, but with a minuscule amplitude that makes it subdominant to other emission channels.
  The dominant secondary electromagnetic mode is superluminal at emission, is subject to weak induced scattering
  within the outflow, and can reach the observer in the radio band.
  \hfil\break \vskip .3in
\end{abstract}

\section{Introduction}\label{s:intro}

The emission of a bright burst of $\sim 10-100$ cm  radiation by a relativistic outflow
from a magnetar poses a challenging problem in multi-scale plasma physics.  This general approach to
the emission of a fast radio burst (FRB) has received significant attention in recent years
(see \citealt{lyubarsky2014,lyubarsky2020,belob2017,plotnikov2019,metzger2019,sironi2021,mahlmann2022}, and
\citealt{lyubarsky2021} for a detailed review).  The radio wavelength 
is a tiny fraction of the width of the outflow, even one lasting for the brief duration of a millisecond.

This paper is based on the observation that the bright X-ray bursts produced by magnetars
provide independent evidence for the emergence of
small-scale structure in the magnetic field.  Energy transfer
to radiating electrons and positrons is mediated by high-wavenumber current perturbations
\citep{t2008,tg2014,nattila2022}.
On occasion, strongly magnetized plasma in this perturbed state may be ejected from a magnetar.
The perturbations are composed of subluminal plasma modes that are highly elongated along the magnetic field
and, therefore, are easily frozen by relativistic expansion.

Here, we investigate the linear interaction of such a frozen Alfv\'en mode or entropy mode with
a caustic (shock) forming in the outflow.  The seed mode is partly converted
to an electromagnetic mode that can escape as a radio wave.
This effect has previously been demonstrated in the case of very rapid plasma 
expansion:  there is efficient linear conversion from a subluminal to a superluminal mode
when the plasma skin depth $c/\omega_p$ expands beyond the size of
the frozen mode \citep{t2017}.    We show that the interaction with
a highly magnetized shock wave has a similar effect:  the mode shrinks compared with
the skin depth as the plasma passes to the downstream side and develops a dynamic electromagnetic component.

This emission channel is shown to be
competitive with a synchrotron maser operating at the same shock (e.g. \citealt{plotnikov2019,sironi2021}),
and will dominate when the magnetization is very high or the
upstream particles are relativistically warm.   The radio wave
naturally has a high degree of linear polarization, which tends to be orthogonal to that produced by the maser.
The outgoing wave amplitude is proportional to the amplitude of frozen turbulence advected with the
relativistic outflow;  the non-linearity is in the background flow.

It is interesting to note
that evidence for such advected structure
in the wind from a magnetically active star comes from a very different, and independent, direction:
measurements of high-wavenumber Alfv\'enic disturbances and current sheets in the Solar wind by the
Parker Solar Probe (e.g. \citealt{bale2019}).  Related structures may also form in other relativistic
outflows from compact stars.

This work is motivated by the detection of two closely spaced radio bursts of luminosity $\sim 10^{37}$ erg s$^{-1}$
from a Galactic magnetar SGR J1935$+$2154 \citep{chime2020,bochenek2020}.
This source produced many X-ray bursts that showed no
detectable radio emission \citep{lin2020a}; the radio-emitting burst was not conspicuously bright but somewhat harder
spectrally than most \citep{younes2021}.

The required small-scale magnetospheric structure 
can be generated by yielding along a fault-like structure
in the magnetar crust, combined with the excitation of small-scale modes by a current-driven
instability in the magnetosphere \citep{td2001,pbh2013,tyo2017,chen2017}.
We posit that radio emission is associated with an electromagnetic explosion that arises when
fault slippage extends close to one of the magnetic poles.

Alternatively, an elastic excitation of the crust can generate an escaping electromagnetic pulse
that is accompanied by a secondary plasmoid instability near the Alfv\'en surface
(see the force-free electrodynamics simulations of \citealt{yuan2020,yuan2022}).
The volume-filling spectrum of magnetic modes that we posit is a more natural
consequence of a current-driven instability driven by volumetric shear.  
Twisting of the magnetic field near the pole is sufficient to generate a $\sim 10^{41}$ erg s$^{-1}$
electromagnetic pulse, as is required in the case of SGR J1935$+$2154.  A giant magnetar flare may involve the
reconnection-driven ejection of a much larger plasmoid \citep{lyutikov2006}; this opens up
the possibility that significantly
brighter radio bursts are produced by the same mechanism we describe, but
in much more energetic events.

\subsection{Related Proposals}

Small-scale electromagnetic fast modes are naturally produced by a maser instability
when $e^\pm$ pairs encounter a forward
shock wave \citep{gallant1992,lyubarsky2014,belob2017,plotnikov2019,metzger2019,sironi2021},
or when a compressive disturbance intersects the strong current sheet
carried out by the pulsar wind \citep{lyubarsky2020,mahlmann2022}.
In contrast with the maser instability \citep{babul2020}, the mechanism described
in this paper operates efficiently when the $e^\pm$ are relativistically warm and the
magnetization (the ratio of Maxwell stress to plasma enthalpy, $\sigma = B^2/4\pi w$)
is large. 

The emission of fast modes by magnetic islands forming at a dynamic current sheet 
has been shown to be a promising mechanism
for generating giant radio pulses \citep{philippov2019}.
The current sheet formed outside the corotating magnetosphere, which has
a negligible guide magnetic field, has been implicated in particular \citep{lyubarsky2020,mahlmann2022}.
We note that the existing simulations of the process must be extrapolated by several orders of magnitude in scale
to accommodate the 6-7 decade separation between the radio wave and the low-frequency
strong electromagnetic wave emitted by a rotating neutron star.
The decay rate of the fast mode power generated at a current sheet
over timescales large compared with the plasma timescale has not yet been determined.

Several numerical simulations have already described the spectrum of
magnetic fluctuations generated in a turbulent, relativistic plasma with a strong guide field
\citep{ripperda2021,chernoglazov2021};
in that case fast mode emission tends to be suppressed on small scales by the elongation
of colliding Alfv\'enic wavepackets.  It is possible that this efficiency is greater when the plasma contains
a mixture of guide fields with opposing signs.

An additional consequence of our analysis is that a strongly magnetized shock is an inefficient
reflector of upstream perturbations.  The flow remains relativistic on both sides of the shock,
which substantially weakens the oppositely directed reflected wave.  Reflection by a very thin and dense plasma shell
can transform spatial structure in the upstream magnetic field to frequency structure in the reflected wave,
but only if the relativistic shell is too thin for a shock to form \citep{t2017}.  This contradicts a
recent claim that the rotational modulation of the magnetic field in a pulsar wind can be transformed to
a radio wave by reflection at a shock \citep{y2022}.

The principal alternative possibility, that some FRBs originate in the corotating magnetospheres
of neutron stars, remains open at present but is subject to controversy (\citealt{belob2021b,qu2022};
see \citealt{lyubarsky2021} for an overview of these models).

\subsection{Summary of Results and Plan of the Paper}

We begin in Section \ref{s:expansion} by reviewing potential sources of high-wavenumber magnetic
modes, including (i) a turbulent cascade triggered by current-driven instability in zones of
strong magnetic shear; 
(ii) distributed magnetic tearing involving the interaction of multiple
tearing surfaces, similar to what is observed during current relaxation in tokamaks;
and (iii) direct injection of small-scale currents in concentrated yielding zones.

Then in Section \ref{s:freeze}
we describe how Alfv\'enic wavepackets and current sheets are stretched out along the toroidal
magnetic field of an escaping electromagnetic pulse.  Modes with non-radial polarization $\delta\bB$
preserve their energy and radial wavevector $k_\perp$
(in the frame of the star) even as the non-radial wavevector $k_\parallel$ decays, thereby freezing
the oscillation.  The frequency of an electromagnetic wave produced by a collision of such a 
frozen mode with a shock is $\omega \sim ck_\perp$.  (The Alfv\'en speed approaches the speed of light $c$
when the energy density is dominated by the electromagnetic field.)
The net result is that a spectrum of frozen modes with size $k_\perp^{-1} \sim 10^{-6}\,c\Delta t$
in an electromagnetic pulse of duration $\Delta t$ carries energy
$(\delta B)^2/8\pi \propto k_\perp^{1-\alpha}$ ($\alpha = 3/2 - 5/3$) and so may seed escaping radiation
with an efficiency $\sim 10^{-3}-10^{-4}$.

Section \ref{s:modes} outlines the pairing of upstream, downstream and reflected modes.  We focus
on modes with comoving wavevector comparable to the inverse skin depth, $\wk \sim \omega_p/c$;
the downstream mode is compressed compared with the downstream skin depth and develops a dynamic
electromagnetic component.  An upstream
frozen Alfv'en mode is polarized perpendicular to the mean magnetic field, and excites the ordinary wave
(O-mode)
downstream of the shock.  The position of the shock is unperturbed to linear order, and the reflected
model is absent.

By contrast, an isobaric mode has magnetic perturbation $\delta\bB$ parallel to the mean magnetic field $\bB$ and therefore
involves finite density and pressure perturbations.  A linear oscillation of the shock is excited and
a fast magnetohydrodynamic wave (extraordinary mode or X-mode) is excited downstream.  The frequency and amplitude
of the reflected X-mode, as measured in the frame of the shock, are limited significantly by the
opposing relativistic motion across the shock.  In general, the fast mode emission by the maser instability
strong dominates the reflected extraordinary wave.

Section \ref{s:coupling} presents the full analysis of linear perturbations of a strongly magnetized
shock.  We begin by reviewing the shock jump conditions, obtaining simple analytic results in the
regime of large magnetization but arbitrary shock strength.  These generalize the analytic results
previously obtained by \cite{kc1984} and \cite{zhang2005}, which apply to a strong shock with arbitrary magnetization.
Then we obtain linear relations between an upstream frozen Alfv\'en mode and a downstream ordinary wave,
and between an upstream isobaric mode and a downstream ordinary wave.

Section \ref{s:collision} generalizes the standard four-zone model of colliding relativistic
shells to the case where both shells have extreme magnetization.
The finite strength of the forward shock is described
in terms of the parameter ${\cal S} = (L_{\rm inner}/L_{\rm outer})^{1/4}$ of the inner and outer shells,
and an analytic two shell model is developed.  The wave amplitudes obtained in Section \ref{s:coupling}
in the frame of the shock are boosted to the frame of the observer.

Section \ref{s:accel} describes the acceleration of relativistic shells with initial magnetization
$\sigma \sim 10^4-10^6$
that pass through an initial fireball phase and estimates the radius and which the passage of forward
and reverse shocks through the shells is completed.   The interaction of the outer shell with a
(much less luminous) rotationally-driven magnetized wind is investigated.  It is shown that
the inhomogeneous acceleration of
the ${\bm E}\times{\bm B}$ frame driven by the internal spreading of an isolated shell plays
an important role in determining this interaction and the deceleration of the outermost luminous shell,
and the distance over which a second shell will interact with it.

In Section \ref{s:conclusions}, the wave amplitudes and frequencies
obtained in Section \ref{s:collision} are evaluated in the context of this collision model.
The output in O-mode photons is shown to easy dominate over the intrinsic X-mode maser emission at the same shock.
We also offer some conjectures about these output parameters will depend on the energy of the
magnetar burst.

Three Appendices detail (i) the derivation of the shock jump at high magnetization, (ii) useful relations
between mode variables for the four relevant plasma eigenmodes, and (iii) the
same relations between mode variables as evaluated in the frame of the shock.

Throughout this paper, the shorthand $X = X_n\times 10^n$ records the value of a quantity
in c.g.s. units.  The Landau excitations of electrons become relativistic in magnetic fields stronger
than $B_{\rm Q} = m_e^2c^3/e\hbar = 4.4\times 10^{13}\,{\rm G}$,
where $m_e c^2$ is the electron rest energy and $e$ the magnitude of the electron charge.

\section{Small-scale Currents in Magnetar Flares}\label{s:expansion}

Magnetars occasionally produce brief bursts of X-rays, most commonly with $\sim 0.1$ s duration
and energy $\lesssim 10^{41}$ erg, but also extending to higher energy and longer duration.
The X-ray spectra of these bursts frequently show a quasi-thermal cutoff above $\sim 40$ keV
(e.g. \citealt{lin2020b,kaneko2021}).
Emission from a dense plasma, which is optically thick to electron-photon scattering
and experiences strong photon splitting in one polarization mode,
provides a concise explanation for this basic feature of the X-ray spectra \citep{td1995};
emission from a more spatially extended and dilute pair plasma tends to have a somewhat higher
peak energy \citep{td2001,belob2021a}.  Bursts of intermediate-to-large energy have smooth
light curves that are modulated by the rotation of the star, pointing to
the presence of a `trapped fireball' that persists beyond the main heating process
\citep{feroci2001}.

Efficient heating of the radiating charges appears to depend on
the presence of high-wavenumber current perturbations.
In effect, magnetar X-ray bursts involve the transfer of magnetic energy from large ($\sim$ km) scales,
where the dynamic plasma behaves similarly to a magnetofluid, to much smaller scales where a fluid description
breaks down.  The fastest channels of energy transfer from the electromagnetic field
to the $e^\pm$ involve either charge starvation of small-scale currents,
or Landau damping of strongly sheared Alfv\'en waves on the particle motion
along $\bB$ \citep{t2008,tg2014,nattila2022}.  Charge starvation sets in when the current demanded of the plasma
exceeds the maximum that can be supplied by conduction, namely
$\delta J > en_\pm c$ in a $e^\pm$ gas of density $n_\pm$.
Shear Alfv\'en waves experience Landau damping off strongly magnetized 
$e^\pm$ when the mode phase speed $v_A \simeq c(1+k_\perp^2c^2/\omega_p^2)^{-1/2}$
drops significantly below the speed of light, at wavenumber $k_\perp \gtrsim \omega_p/c$.
Here $\omega_p$ is the plasma frequency.

Importantly, the damped modes are much smaller than the plasma size, but also much larger than the
microscopic Landau orbital of an electron.\footnote{The electron gyroscale
  $r_g \sim \sigma^{-1/2}c/\omega_p$ in magnetic fields weaker than $B_{\rm Q}$.}

A broad spectrum of current fluctuations is naturally generated by a cascade process.
In the application made in this paper, the wavenumber $|k_\perp|$ of the damped modes  must exceed
the wavenumber of escaping electromagnetic modes.  A concrete example involving the
ejection of a strongly magnetized $e^\pm$ fireball is worked out, demonstrating
that this condition is easily satisfied for radio waves.

\subsection{Forcing of Small-Scale Magnetospheric Currents}

Two related proposals have been made about this cascade process.  The starting point is the injection
of strong, localized shear into the magnetosphere, most naturally
by displacement along extended fault-like structures
\citep{td2001,pbh2013,tyo2017,chen2017}.  Evidence for
concentrated magnetic shear comes from the detection of localized
high-temperature emission in the afterglow of magnetar bursts \citep{kb2017}.  Although
mathematical discontinuities in the crustal strain field imply diverging internal magnetic
shear energy \citep{levin2012}, the thin-shell geometry of the magnetar crust\footnote{Most of the
solid strength of the magnetar crust is concentrated near its base in a layer of sub-km thickness.  Global
stressing of such a thin elastic shell to the point of yielding -- most naturally
by the evolving core magnetic field -- generically produces fault-like structures.}
can support fault-like features with a small but finite width.
A time-dependent elastic-plastic-thermal model shows that a horizontal $\sim 10^{15}$ G
magnetic field thickens the faults but does not suppress them \citep{tyo2017}.

(1) The cascade could be mediated by dynamic current perturbations (Alfv\'en waves) that are
excited by a current-driven instability -- for example, the exchange
reconnection process that has been inferred to drive Alfv\'enic modes in the Solar wind
\citep{bale2022} -- or, alternatively, by small-scale structure in the crustal yielding pattern
(Section \ref{s:fault}).  These modes collide and cascade to high wavenumber
\citep{tb1998,tenbarge2021,nattila2022}.
Here, the collision time is comparable to the mode period.  The conservation of energy flux through
$k$-space implies $(\delta B)^2 \omega =$ constant, where $\omega = c{\bm k}\cdot\hat B = ck_\parallel$
is the mode frequency.  The Alfv\'en waves become increasingly elongated over wavenumber and
one obtains a magnetic spectrum
\be\label{eq:spectrum2}
    \delta B^2 \propto |k_\perp|^{1-\alpha}.
\ee
Here $k_\perp =|\bk\times\hat B|$ is the component of $\bk$ perpendicular to $\bB$.
The index $\alpha$ has been variously estimated to lie in the range $5/3$ \citep{gs1995}
to $3/2$ \citep{boldyrev2005,chernoglazov2021}.  The corresponding spectrum of current perturbations is
\be
\delta J \propto k_\perp^{2/3}-k_\perp^{3/4}.
\ee
In the case of a strong electromagnetic pulse loaded with a quasi-thermal photon-pair gas,
charge starvation is found to set in near the critical wavenumber for Landau damping,
$k_\perp \sim \omega_p/c$ (see Equation (\ref{eq:cs}) below).

(2)  The sheared magnetosphere supporting an inhomogeneous current
is susceptible to relatively slow, small-scale magnetic tearing.
This instability feeds off local extrema in the profile of magnetic twist and involves multiple
interacting tearing surfaces, in close analogy to the anomalous process that redistributes
magnetic twist in a tokamak \citep{white2013,t2022}.  These tearing surfaces
may be spaced by a distance as small as the magnetospheric skin depth.  The instability growth
rate is $s \sim 4\pi \delta J/B$, where the current perturbation
$\delta{\bm J}$ is closely aligned with the magnetic
field.  Large-scale magnetic shear can generate structure on scales down to the skin depth,
as is seen in the structure of the linear eigenmodes.

In the non-relativistic case, explosive small-scale reconnection is seen to be triggered by the collision
of tearing surfaces of opposing signs \citep{ishii02,bierwage05};  related phenomena have been
seen in relativistic kinetic simulations \citep{nalewajko16}.  It is possible that a cascade-like process
develops, now with the quantity $s (\delta B)^2$ conserved in ${\bm k}$-space.  From this, one
deduces a spectrum of current perturbations similar to Equation (\ref{eq:spectrum2}),
$(\delta B)^2 \propto k^{-2/3}$.

The most important feature emerging from both of these processes is the formation of
magnetic perturbations that are (i) elongated along the magnetic field and (ii) carry
energy that decreases relatively slowly with the decreasing size of the
perturbation perpendicular to the magnetic field.  For example, the scaling $\alpha = 3/2$
in Equation (\ref{eq:spectrum2})
implies that perturbations of wavelength $\sim 10$ cm could carry $\sim 10^{-4}-10^{-3}$ of the
energy when a magnetar ejects a relativistically magnetized shell of thickness
$c\Delta t \sim 300$ km.

\subsection{High-Wavenumber Cutoff} 

If the current fluctuations advected out by a strong electromagnetic pulse
are to be a viable seed for escaping radio waves of frequency $\nu$, then their
spectrum must extend at least to a wavenumber $|\bk_\perp| > 2\pi\nu/c$.

To check that constraint is satisfied,
we consider a source zone comprised of a magnetic flux bundle anchored near the
magnetic pole of the star and extending to a radius $r$ which might exceed 10 stellar
radii $R$.  The specific example we consider is an Alfv\'enic cascade.
This dynamic flux bundle eventually breaks open as a result
of the build-up of plasma pressure and continued twisting by the crust -- but not before
non-linear interactions by the Alfv\'en waves have generated a broad power-law spectrum
of modes and heated the embedded pairs sufficiently to generate an optically thick and
quasi-thermal $e^\pm$-photon plasma.

The mode wavenumber $|k_\perp|_{\rm max}$ at which charge starvation sets in can be
expressed in terms of the local magnetic field $B \sim B_p(r/R)^{-3}$ near the top of
the flux bundle, the scattering depth $\tau_{\rm T} \sim n_e\sigma_{\rm T} r$ in the same zone,
and the amplitude of the magnetic fluctuation at the stirring scale $\delta B_0$
(see Equation (87) of \citealt{tg2014}),
\be\label{eq:cs}
\begin{split}
\left({|\bk_\perp|_{\rm max}\, c\over \omega_p}\right)^2 &\sim {\bar\lambda_c\over r}
\left({3\tau_{\rm T}\over 2\alpha_{\rm em}}\right)^{(1+\alpha)/(3-\alpha)}\\
& \times \left[{B(r)\over B_{\rm Q}}\right]^{-4/(3-\alpha)}\left({\delta B_0\over B}\right)^{-2}.
\end{split}
\ee
Here, $B_p$ is the polar magnetic field strength,
$\alpha$ is the spectral index (Equation (\ref{eq:spectrum2})) and $\alpha_{\rm em}
\simeq 1/137$ the fine-structure constant.  We express $\delta B$ in terms of a Poynting
luminosity $L_{\rm P} \sim \delta B_0^2 (\Omega_{\rm P}/4\pi) r^2c$ radiated into a solid angle $\Omega_{\rm P}$
near the top of the flux bundle.
The charge-starvation scale is found to sit close to the skin depth; for a spectral index
$\alpha = 3/2$,
\be
   {|\bk_\perp|_{\rm max} c\over \omega_p} \sim 0.6
   {(\Omega_{\rm P} R_6)^{1/2}\,\tau_{\rm T,1}^{5/6} \over L_{\rm P,42}^{1/2}}
   \left({B_p\over 10\,B_{\rm Q}}\right)^{-1/3} \left({r\over 30\,R}\right)^{3/2}.
\ee

Here, the luminosity has been normalized to a bright Soft Gamma Repeater burst.
The choice of scattering depth, $\tau_{\rm T} \sim 10$, corresponding
to a quasi-thermal plasma with a high compactness $\sigma_T L_{\rm P}/\Omega_{\rm P} m_e c^3r$ and
effective temperature too low ($T_{\rm eff} \lesssim 20$ keV)
for the spectrum to relax to a blackbody distribution \citep{tg2014,belob2021a}.
For example, in this situation,
\be\label{eq:teff}
T_{\rm eff} = \left({L_{\rm P}\over \sigma_{\rm SB}\Omega_{\rm P} r^2}\right)^{1/4} =
5.7\,{L_{\rm P,42}^{1/4}\over \Omega_{\rm P}^{1/4}R_6^{1/2}}\left({r\over 30\,R}\right)^{-1/2}\;{\rm keV}.
\ee
A similar value of $\tau_{\rm T}$ would apply to plasma that experiences a
temporary surge in pair creation due to non-thermal particle acceleration, followed
by a relaxation of heating and passive electron-positron annihilation.

The corresponding plasma frequency is comfortably high enough to seed lower-frequency
radiation at greater distances from the magnetar, where the plasma has been diluted by expansion,
\be
   {\omega_p(r)\over 2\pi} = 6.4\times 10^{12}\,{\tau_{\rm T,1}^{1/2}\over R_6^{1/2}}
   \left({r\over 30\,R}\right)^{-1/2}\;{\rm Hz}.
\ee

\subsection{Direct Injection of Currents at Crustal Faults}\label{s:fault}

Yielding of the magnetar crust provides an interesting direct source of small-scale magnetospheric currents.
The creep rate of a plastically deformed Coulomb solid\footnote{The crust is subjected
  to hydrostatic stress a few orders of magnitude larger than its shear modulus.}
depends even more strongly on the applied stress than it does on temperature \citep{chugunov2010}.
In the presence of small-scale magnetic irregularities (such as may be generated by Hall
drift in the solid crust; \citealt{gourg2022}), the creep rate can vary strongly over small distances.
In this way, small-scale crustal currents may be imprinted in the magnetosphere in the form of
strong cross-field gradients in the magnetic field-aligned current density.

\begin{figure}
\includegraphics[width=\columnwidth]{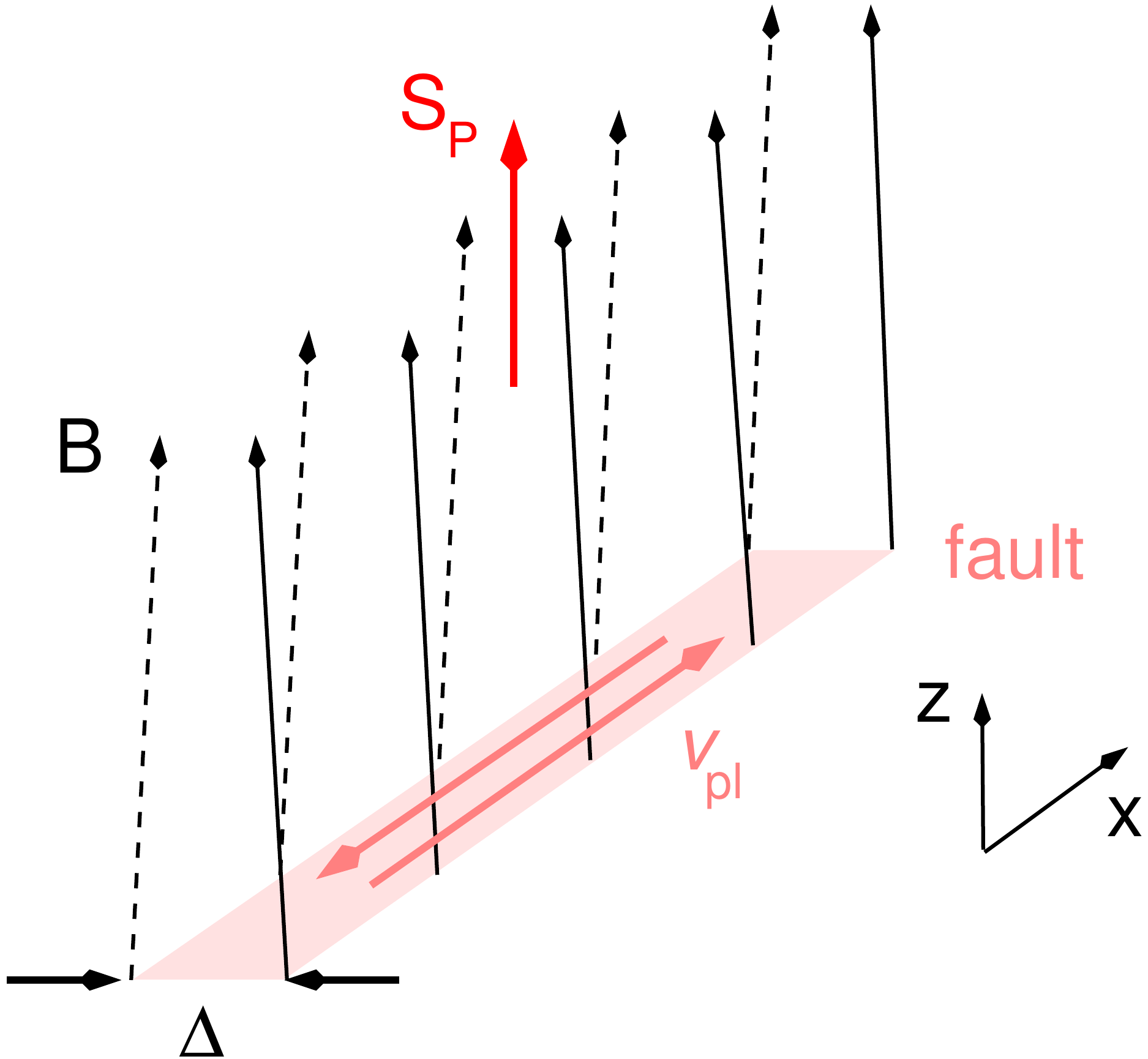}
  \vskip 0.2in
  \caption{Continuous plastic flow with speed $v_{\rm pl}(y)$
    along a fault-like feature of finite width $\Delta \sim 0.1-0.3$ km in the
  magnetar crust.  This flow drives a longitudinal component $B_x$ of the magnetic field, a vertical current
  ${\bm J} = J_z\hat z \sim (B/4\pi)\partial_yv_{\rm pl}\,\hat z$, and a vertical
  Poynting flux ${\bm S}_{\rm P} \sim (v_{\rm pl}^2/4\pi c)B^2\,\hat z$.  Here, the
  background magnetic field is $\bB = B\hat z$ and $B_y = 0$.  The plastic flow rate
  has a strong non-linear dependence on applied stress \citep{chugunov2010}, with the interesting consequence
  that small-scale structure in the crustal Maxwell stress may be imprinted in the flow rate
  and the magnetospheric current.}
\label{fig:fault}
\end{figure}

The $\sim 0.1$ s durations of the most common X-ray bursts emitted by magnetars \citep{gogus2001}
point to an origin in a global disturbance of the magnetar crust:  the burst duration is comparable
to the time for an elastic wave to propagate around the star.  A burst of low energy can
be generated by a limited and localized slippage along a segment of a fault network.  
We can represent this, as in Figure \ref{fig:fault}, as localized plastic flow along a cartesian fault
pointing in the $x$-direction with thickness $\Delta$ in the $y$-direction.  A toy model
of the sub-surface flow can be found in \cite{lander2016}, and a global elastic-plastic-thermal
model of the crustal demonstrating such features in \cite{tyo2017}.

The flow speed ${\bm v}_{\rm pl} = v_{\rm pl}(y)\hat x$ is a function of $y$ and (in the low-energy event
investigated here) is assumed to vanish at $|y| = \Delta/2$.  The background magnetic field is
taken to be vertical and uniform in the $(x,y)$ plane, $\bB = B\hat z$.  The horizontal electric
field at the surface of the star vanishes in the local rest frame of the creeping surface;
hence, ${\bm E} = (v_{\rm pl}/c)B\,\hat y$.  The charge density at the surface is
\be
\rho = {\bnabla\cdot{\bm E}\over 4\pi} = {B\over 4\pi c}\partial_yv_{\rm pl}.
\ee

In the case of the large magnetospheric
current density considered here, this charge density is naturally supplied by
a mild polarization of a collisional and trans-relativistic $e^\pm$ plasma state, as described by
\cite{tk2020}.  The magnetic field lines considered here extend to a large distance from the
star, and so this pair plasma will flow trans-relativistically upward.

A steady-state solution involves a uniform vertical charge flow $J_z = \rho v_{\rm dr}$
moving with drift speed $v_{\rm dr} \sim c$.  This vertical current generates a horizontal magnetic field
\be
B_x = -\int dy {4\pi\over c}J_z = -{v_{\rm pl}v_{\rm dr}\over c^2}B = -{v_{\rm dr}\over c}E_y.
\ee
The vertical Poynting flux is then
\be
S_{\rm P,z} = -{E_yB_x\over 4\pi}c = \left({v_{\rm pl}\over c}\right)^2{B^2\over 4\pi}\,v_{\rm dr}.
\ee
Given that the creep velocity is a fraction $\varepsilon_{\rm pl}$ of the shear wave speed
at the base of the crust ($v_{\rm sh} \simeq 1\times 10^8$ cm s$^{-1}$; \citealt{strohmayer1991}), we obtain
a Poynting luminosity
\be
S_{\rm P,z} \cdot ({\rm km})^2 = 1.3\times 10^{41} \varepsilon_{\rm pl,-1}^2{v_{\rm dr}\over c}
\left({B\over 10\,B_{\rm Q}}\right)^2\quad {\rm erg~s^{-1}}.
\ee
from a patch of crust of area $({\rm km})^2$.   The horizontal creep time is
\be
t_{\rm pl} \sim {\Delta\over\varepsilon_{\rm pl}v_{\rm sh}} \sim 3{\Delta_{4.5}\over\varepsilon_{\rm pl,-1}}
\;{\rm ms}.
\ee

Near one of the magnetic poles, this Poynting flux can flow to a large radius and escape directly.
A second estimate of the escaping luminosity is obtained by 
approximating the surface flow in the polar yielding zone as enhanced rotation with frequency
\be
\Omega_{\rm eff} = {2\pi\rho c\over B} = {1\over 2}\partial_y v_{\rm pl} \sim {v_{\rm pl}\over\Delta}.
\ee
The corresponding luminosity from one hemisphere can be estimated as (following \citealt{spitkovsky2006}),
\be
\begin{split}
L_{\rm P} &\sim \left({1\over 8}-{1\over 4}\right)\left({\Omega_{\rm eff}R\over c}\right)^4B^2R^2c \\
&= (0.9-1.8)\times 10^{43}\,R_6^6\left({B\over 10\,B_{\rm Q}}\right)^2
{\varepsilon_{\rm pl,-1}^4\over\Delta_{4.5}^4} \quad{\rm erg~s^{-1}}.\\
\end{split}
\ee

This estimate is self-consistent as long as the width $\Delta$ of the plastic zone is larger than
the diameter of the field bundle that is opened up by the enhanced current,
\be
\Delta > 0.7\,\varepsilon_{\rm pl,-1}^{1/3}\quad{\rm km}.
\ee
This approach also allows us to estimate the radius at which the electromagnetic field becomes
quasi-transverse and escapes the corotating magnetosphere,
\be
r_0 \sim {c\over\Omega_{\rm eff}} \sim {\Delta c\over v_{\rm pl}} = 1\times 10^8\,
{\Delta_{4.5}\over \varepsilon_{\rm pl,-1}}
\quad{\rm cm}.
\ee

One expects the outgoing Poynting flux to persist at least for the
light-travel time $r_0/c$, because the crustal flow is coordinated over a dimension
$\Delta$ and has a minimum characteristic duration $\Delta/v_{\rm pl}$.
The simplest case is where the radial thickness of the pulse is comparable to $r_0$,
\be
\Delta r = c\Delta t \sim r_0.
\ee
We adopt this simplification in what follows.
Other forcing mechanisms naturally produce pulses of width
$c\Delta t > r_0$;  a nice example is forcing by a periodic elastic oscillation,
as represented in the three-dimensional force-free electrodynamic simulations of \cite{yuan2020,yuan2022}.

As we now discuss,
transverse structure in the current is carried along the magnetic field and will also flow to large distances
from the star.

\begin{figure}
  \includegraphics[width=\columnwidth]{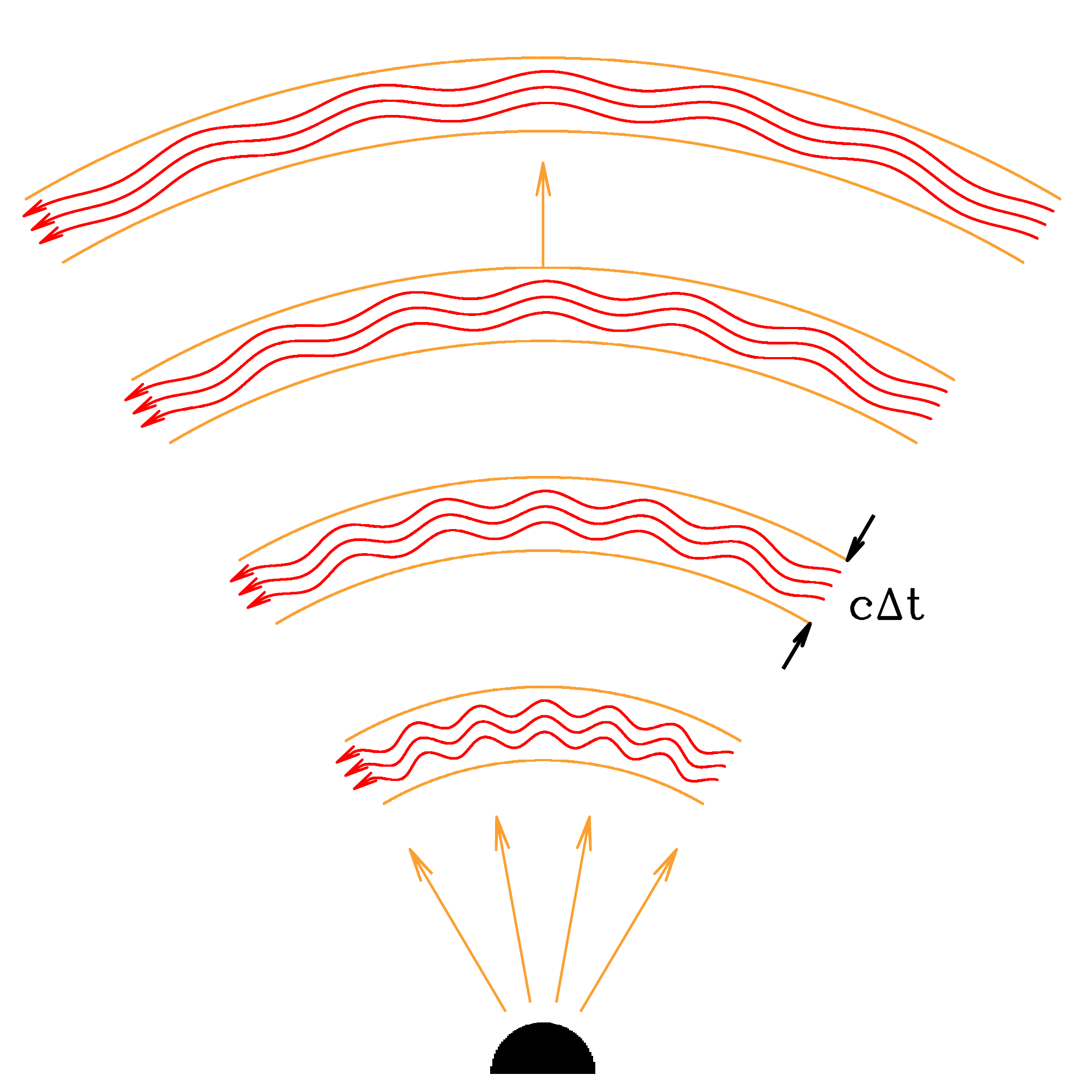}
  \vskip 0.2in
  \includegraphics[width=\columnwidth]{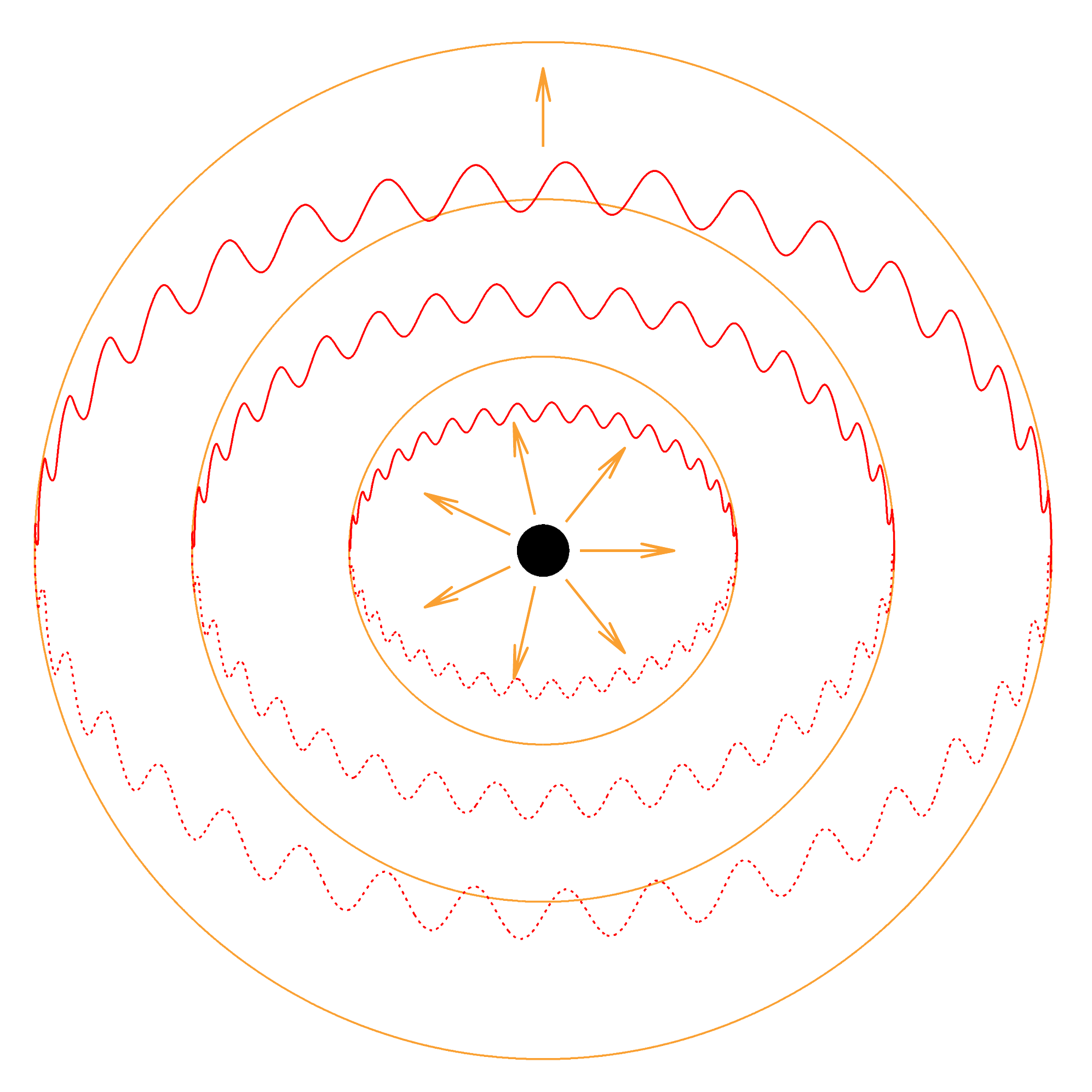}  
\vskip 0.2in
\caption{Transverse electromagnetic modes (Alfv\'en waves) frozen into an expanding relativistic shell
  (of duration $\Delta t$ and thickness $\simeq c\Delta t$).  Only the component of $\bk$
  parallel to the mean magnetic field is depicted.
  Top panel: cross section through a radial sequence of shells.
  Waves polarized in the direction of the flow ($\delta\bB \propto \hat r$) are supported by a radial magnetic field
  that decays more rapidly away from the star than the transverse electromagnetic field
  ($B_r/B_\phi \sim r^{-1}$).   Bottom panel: two-dimensional projection of the magnetic field within
  a sequence of spherical shells.  Waves polarized within the shell
  are stretched in the non-radial direction and therefore dominate the small-scale modes at
  large distances from the star.  These waves are polarized in the plane of a shock formed
  by the self-intersection of the outflow ($\delta\bB \propto \hat\theta$)
  as are the secondary modes produced by their interaction with the shock.}
\label{fig:sphere}
\end{figure}

\section{Freezing of Magnetic Perturbations in an Outflow}\label{s:freeze}

We now turn to consider the effect of relativistic expansion on  small-scale currents that are
imprinted on a large-amplitude hydromagnetic wave.  A relevant example is a 
nearly force-free mode satisfying $\delta\bm J\times\bB = 0$ to linear
order.\footnote{Additional terms in the Lorentz force arising from a background current or electric
field are of secondary importance.}

A current perturbation generated near a magnetic pole of the star will flow outward to a large radius.
Close to the magnetar, one may write $\delta{\bm J} = \delta J_0({\bm x}_\perp, x_\parallel)
e^{i\phi(x_\parallel-ct)}\hat B$.  Here we have separated out the phase variation along $\bB$,
which can be rapid in the case of a dynamic perturbation (a shear Alfv\'en wave).

The shape of the current envelope $\delta{\bm J}_0$ is preserved in the (local) coordinate frame
$\{{\bm x}_\perp\}$ extending transverse to the mean magnetic field $\bB$.  The envelope is stretched
in this transverse plane as the poloidal field lines diverge away from the star,
\be
\begin{split}
\bnabla\times(\delta{\bm J}\times \bB) &=
   (\bB\cdot\bnabla)\delta{\bm J} - (\delta{\bm J}\cdot\bnabla)\bB -(\bnabla\cdot\delta{\bm J})\bB\\
&= {\partial\over\partial x_\parallel}\left({\delta J_0\over B}\right)e^{i\phi}\cdot B^2\hat B = 0.
\end{split}
\ee
Here, $dx_\parallel = \hat B\cdot d{\bm x}$. The terms involving the phase gradient vanish
as a consequence of the conservation of the four-current, involving a finite space charge density $\delta\rho$,
\be
   {\partial(\delta\rho)\over\partial t} + \bnabla\cdot\delta{\bm J} = 0;  \quad
   \delta\rho = {1\over c}\delta J_0e^{i\phi}.
   \ee

The forcing of the magnetosphere by crustal motions persists for a limited interval $\Delta t$;
the background field (on which small-scale irregularities are superposed)
transforms at a distance $r_0 \sim c\Delta t \sim 300\,(\Delta t/{\rm ms})$ km
to a large-amplitude electromagnetic wave.  This wave propagates outward
subluminally; at $r > r_0$ there is a frame moving with radial speed $\beta_{E\times B}c$
in which the electric field ${\bm E}$ nearly vanishes.
The Lorentz factor $\Gamma = (1-\beta_{E\times B}^2)^{-1/2}$ grows with radius close to the star,
$\Gamma(r) \sim r/r_0$.  In what follows, this relativistic expansion will be approximated as locally spherical,
with mean magnetic field $\bB = B\hat\phi$.

Now consider how a current perturbation responds to this expansion.  Its
amplitude and scale $\ell$ in the direction transverse to $\bB$ are of particular interest,
because these quantities determine ultimately the amplitude and wavelength of the escaping radiation.
At the base of the wave zone, the degree of elongation along the magnetic field depends on whether the
perturbation is generated close to the star, or by a more distributed process operating near radius $r_0$
(e.g., turbulent cascade or magnetic tearing).  In the first case, the width increases to
\be
\ell_0 \equiv \ell(r_0) \sim \left({c\Delta t\over R}\right)^{3/2}\ell(R)
\ee
from a value $\ell(R)$ at the surface of the star (radius $R$).

Moving next to the relativistic expansion phase, we will generally work in a frame comoving
with the flow.
Note that the current perturbation $\delta{\bm J}$ is aligned with the toroidal magnetic field,
but ${\bm k}_\perp$ is only
constrained to lie in a plane perpendicular to $\bB$.  We may consider how a mode responds
to the expansion in the idealized case where it is freely propagating and does not experience
collisions with other modes.  When ${\bm k}_\perp$ is aligned
with the direction of the flow, then $\ell = |\bk_\perp|^{-1}$ is conserved in the frame
of the star, but expands as
\be\label{eq:lexp}
\ell(r) \sim \Gamma(r)\,\ell_0;  \quad\quad (\bk\,\parallel\,\hat r)
\ee
in the comoving frame.
Modes with non-radial ${\bm k}_\perp$ are stretched according to
\be\label{eq:lexp2}
\ell(r) \sim {r\over r_0}\ell_0.  \quad\quad (\bk\,\parallel\,\hat \theta)
\ee

We next summarize a few important qualitative effects of expansion on embedded current perturbations.

(1) The current perturbations generally have a finite frequency of oscillation
but these oscillations slow down as the irregularities are stretched out in the expansion
(Figure \ref{fig:sphere}).   A structure in the magnetic field extending to an angle
$\Delta\phi > 1/\Gamma$ becomes frozen.
For example, an Alfv\'en mode has a frequency
$\omega = c{\bm k}\cdot \hat B \simeq k_\phi c$ at large magnetization;
in the present context the wavepacket is elongated
along the mean magnetic field, i.e., $k_\phi \ll |{\bm k}| \sim 1/\ell$.
The non-radial wavenumber redshifts as
\be\label{eq:kexp}
k_\phi(r) = \left({r_0\over r}\right) k_{\phi0},
\ee
in the relativistic expansion phase; here $k_{\phi0} = k_\phi(r_0)$ is the wavenumber
at the base of the outflow.

One notes that expansion preserves the shape of a wavepacket
in the inner part of the expansion, where $\Gamma \propto r$, but increases its
elongation as the bulk acceleration slows.
During the first 
  phase, the expansion is nearly isotropic in the comoving frame:  two closely separated
  points pull apart at a rate
\be
  {d\ln(\Delta\ell)\over dt} = {d\ln(r\Delta\theta)\over dt} = {d\ln(r\Delta\phi)\over dt} \sim {c\over r_0}.
\ee
The time
lapsed in the comoving frame increases only logarithmically with radius, $dt/\ln r
\sim r/\Gamma c \sim r_0/c$.   In analogy with
an inflationary phase of cosmic expansion, perturbations are pulled outside the causal
horizon by the expansion.

An increasing portion of the mode spectrum is frozen as the flow expands.
Alfv\'en modes stop oscillating when their non-radial wavenumber drops below
\be\label{eq:kcrit}
k_\phi < k_\phi^F(r) = {\Gamma\over r}.
\ee
It is convenient to measure this effect in the Lagrangian space $k_{\phi0}$,
where the critical wavenumber is
\be\label{eq:kcrit2}
k_{\phi0}^F(r) = {\Gamma\over r_0}.
\ee
Modes with a higher wavenumber are constantly regenerated by a cascade
that starts at an effective stirring scale given by Equation (\ref{eq:kcrit})
or (\ref{eq:kcrit2}).  This stirring scale progressively shrinks (in Lagrangian
wavenumber) as the flow expands.

(2) The wavepacket thickness can expand dramatically compared with the electron
skin depth $d = c/\omega_p$.  If embedded $e^\pm$ left over from a quasi-thermal fireball
experience only adiabatic cooling, they will cool to sub-relativistic
temperatures. Their comoving density is related to the Thomson
depth $\tau_{\rm T,0} \sim 10$ at the launching radius by
\be\label{eq:nvsr}
n_\pm = {\tau_{\rm T,0}r_0\over \Gamma(r)\sigma_{\rm T} r^2}.
\ee
The plasma frequency $\omega_p = (4\pi ne^2/m_e)^{1/2}$ decreases as
$\omega_p \sim \Gamma^{-1/2} r^{-1}$.

We infer that the current
gradient scale shrinks compared with the expanding skin depth,
\be
   {\ell\over d} \propto {\Gamma^{1/2}\over r}.
\ee
A change in the relative size of $\ell$ and $d$ can be associated with a transition from
a subluminal to a superluminal mode \citep{t2017}.

(3) Shocks forming in the outflow can modulate the properties of advected
magnetic modes in similar ways.
In this paper, we demonstrate how superluminal electromagnetic waves may be excited
by the collision of frozen magnetic perturbations with shocks.
The skin depth $d$ hardly changes downstream of the shock,
whereas the comoving mode wavelength $\ell$
shrinks by a factor $\sim \gamma_2/\gamma_1$, with the result that $\ell/d$ decreases.
(Here, $\gamma_1$ and $\gamma_2$ are the upstream and downstream flow Lorentz
factors in the frame of the shock; see Section \ref{s:jump} and Appendix \ref{s:shock} for further details.)

\subsection{Preferred Orientation of the Frozen Modes}

The stretching of frozen magnetic irregularities depends on their orientation
(Figure \ref{fig:sphere}).

We first consider transverse perturbations with $\delta\bB \perp \bB$
and $\delta{\bm J} \parallel \bB$.  
It is easy to see that, after some expansion, the dominant advected mode is
$\delta{\bm B} = \delta B_\theta\,\hat\theta$ (as measured in local spherical coordinates)
with the gradient pointing in the radial direction.  The wave is frozen into the expanding flow and so
$\delta B \sim \delta B_\theta \propto (\Gamma r)^{-1}$ and $\ell \propto \Gamma$ in the comoving frame.
The fraction of the outflow energy carried by this polarization is invariant under expansion.
The current perturbation scales as $\delta J_\phi \propto (\Gamma^2 r)^{-1}$.
In comparison, the radial magnetic perturbation with non-radial gradient decays as $\delta B
\sim \delta B_r \propto r^{-2}$, $\ell \propto r$, and $\delta J_\phi \propto r^{-3}$.

These two scalings coincide only when the outflow expands rapidly as
$\Gamma \propto r$ -- as it does inside the fast magnetosonic point. 
During the later stages of the expansion, the increase of $\Gamma$ is generally slower.
We conclude that the non-radial magnetic perturbations dominate at large radius.
These modes are polarized in the plane of a shock formed by caustic in the outflow --
as are the secondary modes formed by the interaction with the shock, which can eventually
escape as radio waves.  Our study of mode-shock interaction is restricted to this case.

Consider next an isobaric mode with finite pressure perturbation, $\delta P = -B_\phi\,\delta B_\phi/4\pi$.
In the case where the pressure is supplied by relativistic $e^\pm$, the pressure perturbation
evolves differently under adiabatic expansion from the magnetic pressure perturbation,
as $\delta P \propto n_\pm^{4/3} \propto (r^2\Gamma)^{-4/3}$ in comparison with
$B_\phi\delta B_\phi \propto B_\phi^2 \propto (\Gamma r)^{-2}$.  A static isobaric mode is
therefore converted to a dynamic fast mode, which may be damped.  In addition, the high
magnetization of a quasi-thermal fireball at its emission radius $r_0 \sim c\Delta t$
implies a hard upper limit to the amplitude of the seed
isobaric mode as compared with a seed Alfv\'en mode:  $(\delta B_\phi/B_\phi)^2 \sim
\sigma_0^{-2} \sim 10^{-8}$ for a burst of energy $\sim 10^{39}$ erg, duration $\Delta t \sim 10^{-3}$ s,
and initial Thomson optical depth $\tau_T \sim 10$.  (See the fireball model outlined in
Section \ref{s:accel_single}.)

\subsection{Width and Energy of the Frozen Modes}\label{s:cutoff}

The escaping electromagnetic signal that we calculate in Sections \ref{s:accel} and \ref{s:conclusions}
is most sensitive to
the radial wavenumber and energy of the frozen modes.  The distribution of shear Alfv\'en waves
in wavenumber space $\{\bk_\perp,k_\parallel\}$ is found by
combining the magnetic power spectrum (Equation (\ref{eq:spectrum2})) with the constraint of
a conserved energy flux, $\omega \delta B^2 = ck_\parallel \delta B^2 =$ const.  One obtains
the simple scaling
\be\label{eq:spectrum3}
\ell(k_\phi) \sim \ell_{\rm stir} \left({k_\phi\over k_{\phi,\rm stir}}\right)^{-1/(\alpha-1)},
\ee
where $\alpha = 3/2-5/3$.

So far we have considered the effect of expansion on free modes.  A next consideration
is its effect on the strength of the coupling between oppositely propagating Alfv\'en modes.
We first consider the initial expansion, where $\Gamma(r) \sim r/r_0$.
  Then the two
  polarization states evolve similarly in the comoving frame, and we consider the single
  parameter $\delta B_\theta/(k_\phi \ell)B_\phi$ \citep{gs1995}.
We evaluate this in Lagrangian space, at fixed wavenumber $k_{\phi0}$.  Then
$\delta B_\theta/B_\phi$ is not changed by expansion and the scalings (\ref{eq:lexp}) and
(\ref{eq:kexp}) imply
\be
   {\delta B_\theta\over (k_\phi \ell) B_\phi}
   \sim {r/r_0\over\Gamma r_0}{\delta B_\theta\over (k_\phi \ell) B_\phi}\biggr|_{r_0} \sim {\rm const}.
\ee
Mode collisions are suppressed only by the finite expansion time.

As the hydromagnetic flow expands, the outer (`stirring') scale of the mode spectrum
shrinks in Lagrangian space.  The radial width of a mode of fixed $k_{\phi0}$ is invariant in the
frame of the observer, $\ell_{\rm obs} = \ell/\Gamma = \ell_0$, and determines
the wavelength of the escaping radiation.  We therefore wish to determine the energy
carried by frozen modes at a given $\ell_0$ (or $k_{\phi0}$).  Modes of wavenumber above
the stirring scale (\ref{eq:kcrit2}) are continuously regenerated.  Therefore the
energy spectrum of the frozen modes, as evaluated in Lagrangian space, is essentially
the same as the spectrum (\ref{eq:spectrum2}) of turbulence in a static box.

The minimum radial width $\ell_0^F$ of the frozen modes is obtained by making use of Equation (\ref{eq:spectrum3})
in Lagrangian space and taking $k_{\phi,\rm stir} \sim 1/r_0$ at the base of the outflow.
Therefore,
\be\label{eq:ell0F}
   (\ell_0^F)_{\rm min}
   \sim  (r_0 k_{\phi 0}^F)^{-1/(\alpha-1)}\ell_{\rm stir}(r_0)
   \sim {\ell_{\rm stir}(r_0)\over \Gamma^{1/(\alpha-1)}},
\ee
where we have substituted Equation (\ref{eq:kcrit2}).

To evaluate the strength of mode collisions farther out in the expansion, we must
  consider polarization effects in more detail.  Now the growth in $\Gamma(r)$
  has slowed from the initial linear growth phase.  The fraction of the 
  energy carried by the radially polarized modes decreases in the absence of
  collisions, $\delta B_r^2/\delta B_\theta^2 \sim [\Gamma /(r/r_0)]^2$.
  Both wave polarizations may be decomposed into oppositely propagating Elsasser modes,
which are
\be
w_{r,\theta}^\pm \simeq 2{\delta B_{r,\theta}^\pm\over B_\phi}
\ee
at high magnetization.  Here, mode $+$ ($-$) propagates to larger (smaller) $\phi$. The effect of a collision
between modes $\mp$ and $\pm$ on the propagation of modes $\pm$ may be written as (e.g. \citealt{lg2007})
\be\label{eq:elsass}
 \Delta{\bm w}^\pm \sim {({\bm w}^\mp\cdot\bnabla){\bm w}^{\pm}\over ik_\phi} \;\sim\;
   \left[\left({\delta B^\mp_\theta k_\theta\over k_\phi B_\phi}\right)w^{\pm}_r,\;\;
   \left({\delta B^\mp_r k_r\over k_\phi B_\phi}\right)w^{\pm}_\theta\right].
\ee
One sees that an Alfv\'en mode with a given polarization is modified by colliding with Alfv\'en modes with
     the orthogonal polarization.  Substituting the radial scalings for $\delta\bB$ and $\bk$
     derived previously, one finds that the coupling parameter multiplying either component of ${\bm w}^{\pm}$
     in Equation (\ref{eq:elsass}) is invariant under the effects of expansion. 

     Consider, in particular, the case where a part of the outflow has stopped accelerating outward,
     $\Gamma(r) \sim$ const.  The expansion is now anisotropic in the comoving frame:
     $d\ln(\Delta l)/dt = 0$ whereas $d\ln(r\Delta\theta)/dt \sim d\ln(r\Delta\phi)/dt \sim 1/t$.
     Modes with $k_\phi \sim k_{\phi,\rm stir} \gtrsim
     \Gamma/r \sim 1/ct$ experience only a few oscillations.
     (This situation corresponds to an expanding universe with nearly constant comoving horizon size.)
     These modes have fixed
     comoving wavenumber $k_{\phi 0} \sim \Gamma/r_0 \sim {\rm const}$ and radial with $\ell_0^F$ given
     by Equation (\ref{eq:ell0F}).
     Equation (\ref{eq:elsass}) implies that these oscillating modes transfer a significant part of their
     energy to higher wavenumbers.  The rate of energy
     deposition by this cascade -- normalized by the total magnetic energy density $B_\phi^2/8\pi$ --
     now scales as $1/t$.
   
Given that a part of the outflow reaches a limiting Lorentz factor $\Gamma_{\rm max}$, one infers
\be
(\ell_0^F)_{\rm min} \sim {\ell_{\rm stir}(r_0)\over \Gamma_{\rm max}^{1/(\alpha-1)}} <
   {c\Delta t\over\Gamma_{\rm max}^{1/(\alpha-1)}}.
\ee
For a spectral index $\alpha = 3/2$, this implies a wavelength at least as short as
$c\Delta t/\Gamma_{\rm max}^2$.  Inverting this relation, one finds a minimum Lorentz
factor that is needed to freeze modes of radial wavenumber $k_\perp$ and frequency $\omega = ck_\perp$,
as observed in the frame of the star:
\be
\begin{split}
\Gamma_{\rm max} &> (\omega \Delta t)^{\alpha-1} \\
&\sim (\omega\Delta t)^{1/2} = 2.5\times 10^3\,\nu_9 (\Delta t_{-3})^{1/2}.\quad (\alpha = 3/2)\\
\end{split}
\ee

\begin{figure}
\includegraphics[width=\columnwidth]{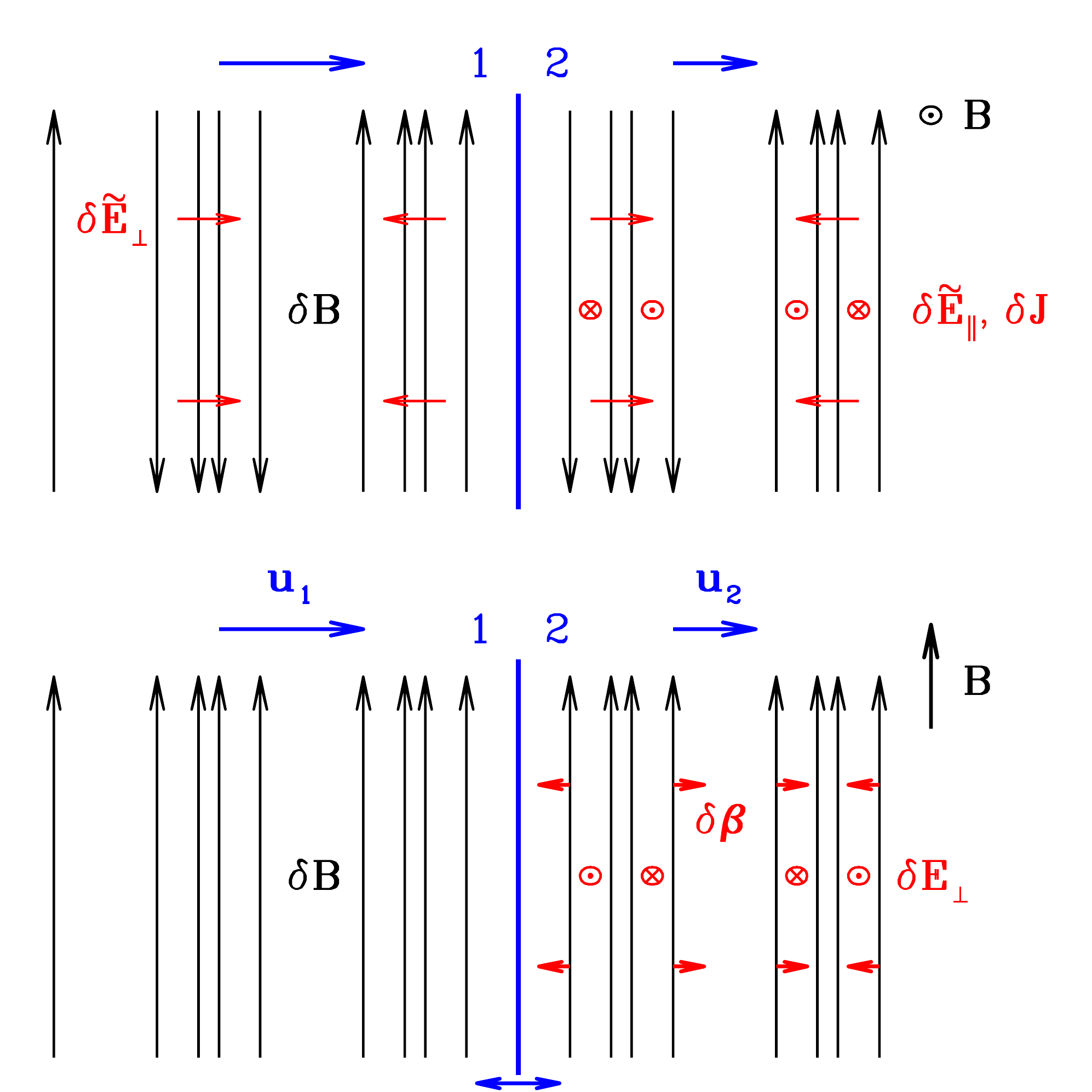}  
    \vskip 0.2in
\caption{Linear perturbations of a shock in a relativistically magnetized plasma flow with field $\bB$ oriented in the plane of the shock.  Upstream Lorentz factor $\gamma_1 > \gamma_{\phi,\rm X}(\sigma_1) =  \sqrt{3\sigma_1/2} \gg 1$ (see Equation (\ref{eq:gamphiX})).  
  The flow is perturbed on the upstream side by a mode with comoving frequency $\womega = 0$
  and wavenumber $\wk_1 \sim \omega_{p1}/c$.  In the frame of the shock, upstream and downstream flow variables oscillate
  with the common frequency $\omega = \beta_1\gamma_1 c\wk_1$.   A second mode with finite comoving phase speed is excited downstream
  of the shock.   Top panel: upstream mode is a frozen Alfv\'en wave polarized $\delta\bB \perp \bB$; the downstream excitation includes
  an ordinary electromagnetic wave (O-mode).  Bottom panel:
  upstream mode is a compressive isobaric mode with $\delta\bB \parallel \bB$ and pressure perturbation
  $\delta P  = - B\delta B/4\pi$;  the downstream excitation includes the X-mode (fast magnetosonic mode).
  The shock position
  is perturbed to linear order when the upstream mode is compressible (bottom panel).  We focus on
  ideal modes with wavelength much larger than the $e^\pm$ gyro-radius;  independent plasma simulations indicate
  that a maser-driven instability of high-frequency fast waves is present at the shock, but with small amplitude at large
  magnetization $\sigma_1$.
  A weak X-mode is reflected into the upstream flow, but only if $c\wk_1 \lesssim
  \omega_{p1}/\gamma_1$.}
\label{fig:shock}
\end{figure}

\section{Secondary Wave Modes Excited Near a Shock}\label{s:modes}

We describe in this Section four modes that may be excited near a relativistically magnetized shock, and outline
how primary modes frozen into the upstream flow will excite secondary modes on both sides of the
shock (Figure \ref{fig:shock}).  The upstream modes will be taken to have vanishing frequency $\womega$ as measured in the frame
of the plasma flow;  we consider a frozen shear Alfv\'en wave and an isobaric mode.
The primary mode oscillates at a finite frequency in the shock frame,
\be
\omega = \omega_1 = \gamma_1 c\bbeta_1\cdot\wbk_{\rm seed},
\ee
where $\gamma$ is the mean flow Lorentz factor.
The secondary modes oscillate at the common frequency $\omega$ but may also have a propagating character
in the plasma frame.  In what follows, 1 and 2 denote the upstream and downstream sides of the shock.
When a quantity is being considered in both the shock frame and the
plasma rest frame, a tilde will denote its value in the rest frame,
e.g., $\wbB$, $\wbk$, $\wom$ for the comoving flux density, wavevector, and frequency.\footnote{The
  magnetization $\sigma$, plasma frequency $\omega_p$, particle density $n$, plasma
pressure $P$ and enthalpy density $w$, sound speed $c_s$, and effective mass ${\cal M}$ are always defined 
in the plasma frame.}

The magnetic field observed in shock frame
is aberrated into the plane of the shock (the $y-z$ plane) by the relativistic upstream motion.
The wavefront of any upstream perturbation is also aberrated into the shock plane.
Downstream of a strong shock, the comoving flux density is increased,
by a factor $\wB_2/\wB_1 \sim \gamma_1/\gamma_2$, even while the normal component of ${\bm B}$
is conserved.  Hence, one may assume that the comoving magnetic field lies in the plane of the shock
on the downstream side.\footnote{The upstream comoving field may not generally have this orientation;
nonetheless, the aberration effect in the shock frame makes this irrelevant to our conclusions.}

We will therefore focus on the simplest case where the perturbation wavevector is aligned with
the shock normal, $\bk = k\,\hat x$, and choose the background magnetic field
$\bB = B\,\hat y \perp \bk$ on both sides of the shock.

Two additional finite-frequency modes -- the ordinary and extraordinary electromagnetic modes, O-mode
and X-mode -- 
may be excited on the downstream side.  A compressive seed perturbation like the isobaric mode
will also excite a small-amplitude oscillation of the shock.
An interesting case is where the comoving wavevector $\wk_{\rm seed} \sim 1/d_1$.
When the shock is strong, the downstream modes then have a relatively short wavelength,
$\wk_2 > 1/d_2$, enhancing their electromagnetic character.  In this case,
the shock oscillation is too rapid to excite a backward-propagating X-mode wave.

\subsection{Pairing of Upstream and Downstream Modes}

An oscillating electric field is excited in the downstream plasma, driven by (i) the postshock imbalance
between the current perturbation $\delta{\bm J}$ and the magnetic curl $i(c/4\pi){\wbk}\times\delta\wbB$
and, in the case of a compressive seed perturbation, by (ii) an oscillation of the shock.

A downstream X-mode or O-mode wave has a finite frequency in the plasma frame, with a phase
speed $\wbeta_\phi = \womega/c\wk = O(1)$.  The frequency in the shock frame is
\be
\omega_2 = \gamma_2 \left(\womega_2 + \beta_2 c\wk_2\right) = \gamma_2\left(\wbeta_{\phi,2} + \beta_2\right)c\wk_2.
\ee
The downstream perturbation is compressed along with the background magnetofluid, $\wk \propto \wB,\;n$;
it therefore shrinks compared with the electron skin depth $d$.  In the case of a strong shock,
\be
\wk_2 d_2  \simeq 2^{-3/2}\left({\gamma_1\over \gamma_2}\right)\,\wk_{\rm seed} d_1.
\ee
Here, we have substituted Equation (\ref{eq:ompratio}).

The mode excited downstream of the shock depends on the polarization of the upstream perturbation.

(1)  When the upstream magnetic perturbation is transverse to $\wbB$, as it is in the frozen Alfv\'en mode,
the O-mode is excited downstream of the shock,
\be
A[\womega_1 \rightarrow 0] \;\xrightarrow{{\rm shock}}\; A[\womega_2 \rightarrow 0] + O[\womega_2 > \omega_{p2}],
\ee
with comoving frequency $\wom_2 > \omega_{p2}$ (see Equation (\ref{eq:odisp})).
The electric vector of the O-mode is aligned with the background magnetic field $\delta\bE \parallel \bB$.
The O-mode is primarily electromagnetic when $\wk_2d_2 > 1$, with a phase speed approaching the speed of light,
$\wbeta_{\phi2} \rightarrow 1$.
Downstream of a strong shock, a negligible component of the downstream perturbation remains in the A-mode.

(2)  An isobaric seed perturbation 
converts on the downstream side to a weaker isobaric mode
and an oscillating magnetosonic compression (the X-mode) with
electric vector $\delta\bE \perp \bB$ and phase speed $\wbeta_\phi$ given by Equation (\ref{eq:betax}),
\be
I[\womega_1 = 0] \;\xrightarrow{{\rm shock}}\; I[\womega_2 = 0] \;+\; X[\womega_2].
\ee

\subsection{Reflected X-mode}

A compressive excitation of the shock may also excite an X-mode propagating counter to the
upstream flow, but only if the incoming perturbation has a long wavelength, $\wk_{\rm seed} d_1 \ll 1$.
This reflected mode shares the same frequency as the incoming isobaric mode
as measured in the frame of the shock.  An upper bound to the frequency of the reflected
mode is given by the intrinsic maser instability.
The maser-generated mode has essentially the same frequency
as the (comoving) gyro-frequency of particles on the downstream side of the shock; in the case of a strong shock,
one has in the frame of the downstream flow\footnote{The normalization here is close to that
  found numerically \citep{plotnikov2019,sironi2021}; we have expressed the effective mass ${\cal M} = w/nc^2$
  in terms of the plasma enthalpy $w$ and made use of the jump condition (\ref{eq:jump}).}
\be
(\omega_{\rm maser})_2 \sim {eB_2\over \gamma_2{\cal M}_2 c} \simeq 2{eB_1\over \gamma_1{\cal M}_1 c}
= 2\omega_{c1}.
\ee
The comoving cyclotron and plasma frequencies are related by
\be
   {\omega_c^2\over\omega_p^2} = \sigma = {B^2\over 4\pi \gamma^2 w}.
\ee
The frequency of the maser-generated mode in the frame of the shock is
\be
\omega_{\rm maser} = \gamma_2[1-\beta_2/(\beta_{\phi,\rm maser})_2](\omega_{\rm maser})_2,
\ee
where $(\beta_{\phi,\rm maser})_2$ is its phase speed in the frame
of the downstream flow.   We require $(\beta_{\phi,\rm maser})_2 > \beta_2$ for this mode to
propagate upstream of the shock;  hence 
\be
\omega_{\rm maser} \lesssim {\omega_{c1}\over\gamma_2} = {\sigma_1^{1/2}\over\gamma_2}\omega_{p1} \sim \omega_{p1}.
\ee

A reflected X-mode is present only if the seed perturbation has frequency $\omega < \omega_{\rm maser}$
in the frame of the shock.  Hence,
\be\label{eq:X1exists}
  \wk_{\rm seed} d_1 \lesssim {1\over \gamma_1}. \quad\quad (\delta B_{1,\rm X} \neq 0)
\ee
The seed perturbations we consider typically have a higher frequency, in which case
the reflected X-mode is frozen out.  Even when Equation (\ref{eq:X1exists}) is satisfied, we will
see that the reflected mode is subdominant both to the downstream modes and to the maser-generated
upstream mode.

\subsection{Linear Plasma Modes}\label{s:lin_modes}

The polarization, dispersion relation, and relations between the mode variables are summarized here
in more detail.   (We work in the plasma frame and
therefore drop the tilde label in this Section.)  Derivations can be found in Appendix \ref{s:modes_app}.
The perturbation variables are reconstructed in the frame of the shock in Section \ref{s:matching}.

We consider a superposition of zero-frequency modes on the upstream side of the shock.

(1) A frozen Alfv\'en wave with vanishingly small frequency
and magnetic perturbation $\delta\bB \perp \bB$.
The wavevector of this mode is $\bk =  k(\hat x + \varepsilon\hat y)$.
We are interested in the case where the mode
is extremely elongated along the magnetic field, $\varepsilon \rightarrow 0$.
Its frequency vanishes as 
\be
\omega = \beta_{\rm A}c k_\parallel = \varepsilon\cdot\beta_{\rm A} c k,
\ee
where $\beta_{\rm A} = (1+1/\sigma)^{-1/2}$ is the Alfv\'en speed in units of the speed of light.

The polarization of interest
is the one which couples to a superluminal electromagnetic wave on the downstream side
of the shock:
\be
\delta\bB_{\rm A} = \delta B_{\rm A}\,\hat z = \delta B_0e^{i\bk\cdot\bx}\,\hat z;\quad
\ee
The electric vector is also polarized perpendicular to ${\bm B}$,
but has a dominant longitudinal component.

(2) An isobaric mode with $\delta\bB$ parallel to $\bB$.  The perturbation to the Lorentz
force is compensated by a plasma pressure gradient, hence
\be\label{eq:isobar}
\delta P_{\rm I} = -{B\delta B_{\rm I}\over 4\pi}.
\ee
In thermal equilibrium, as assumed here, the plasma state is defined by two thermodynamic variables.
The perturbation is taken to be isothermal upstream of the shock but that assumption cannot be made
on the downstream side.

At high magnetization, the linear isobaric mode is restricted to a much lower amplitude than is the
frozen Alfv\'en mode.  Requiring that $\delta P/P \lesssim 1$ implies that
\be
   {\delta B_{\rm I}^2\over B^2} \lesssim {1\over (4\sigma)^2};
\ee
by contrast, a linear Alfv\'en mode is limited to $\delta B_{\rm A} \lesssim B$.

Two finite-frequency modes are excited on the downstream side.  Both are transverse modes,
in the sense that $\delta\bB,\;\delta\bE \perp \bk$.  

(3) The electromagnetic O-mode has electric vector aligned with $\bB$,
\be\label{eq:emo}
\begin{split}
\delta\bE_{\rm O} &= \delta E_{\rm O}\,\hat y = \beta_{\phi,\rm O}\delta B_0e^{i(kx-\omega t)};\\
\delta\bB_{\rm O} &= \delta B_{\rm O}\,\hat z = -\hat x\times {\delta\bE_{\rm O}\over\beta_{\phi,\rm O}}
\end{split}
\ee
and dispersion relation
\be\label{eq:odisp}
\omega^2 = c^2 k^2 + \omega_p^2.
\ee
This mode is superluminal, with phase speed
\be\label{eq:betao}
\beta_{\phi,\rm O} = {\omega\over c k} = \sqrt{1 + {\omega_p^2\over c^2 k^2}}
\ee
and group speed
\be\label{eq:betago}
\beta_{g,\rm O} = {1\over\beta_{\phi,\rm O}} = {ck\over\sqrt{c^2k^2 + \omega_p^2}}.
\ee
The plasma temperature is relativistic on the downstream side of the shock, and the plasma
frequency is given by
\be\label{eq:omp}
\omega_p^2 = {4\pi e^2 n\over {\cal M}},
\ee
where
\be\label{eq:meff}
   {\cal M} = {w\over nc^2}
\ee
is the effective mass and $w$ is the comoving enthalpy density including
rest energy.

(4) The electromagnetic X-mode excited on the downstream side of the shock has a 
comoving frequency $\ll \omega_{c2}$ and has a hydromagnetic description as the
fast mode.  The magnetic perturbation $\delta\bB$ is aligned with the background field,
\be\label{eq:emx}
\begin{split}
\delta\bB_{\rm X} &= \delta B_{\rm X}\,\hat y = \delta B_0e^{i(kx-\omega t)};\\
\delta\bE_{\rm X} &= \delta E_{\rm X}\,\hat z = \beta_{\phi,\rm X}\,\hat x\times\delta\bB_{\rm X},
\end{split}
\ee
and the mode dispersion relation is
\be\label{eq:betax}
\begin{split}
\beta_{\phi,\rm X} &= \pm\sqrt{\sigma +c_s^2/c^2\over \sigma + 1}\\
&\simeq \pm\left[1-{1-c_s^2/c^2\over 2\sigma}\right].\quad\quad(\sigma \gg 1)
\end{split}
\ee
The corresponding phase Lorentz factor is
\be\label{eq:gamphiX}
\gamma_{\phi,\rm X} = {1\over\sqrt{1-\beta_{\phi,\rm X}^2}} = \sqrt{3(\sigma+1)\over 2}
\ee
when the plasma is relativistically hot and $c_s \simeq c/\sqrt{3}$.
The group speed $\beta_{g,\rm X} \simeq \beta_{\phi,\rm X}$.

The frozen Alfv\'en mode is incompressible and so couples uniquely
to the ordinary electromagnetic mode downstream of the shock.   The isobaric mode couples to a linear
combination of a compressive fast mode and a more complicated isobaric mode (involving a finite
temperature perturbation).   

(5)  A reflected X-mode is present when an isobaric perturbation of wavenumber smaller than
(\ref{eq:X1exists}) collides with the shock.  
The nature of this reflected mode depends on whether its frequency in the frame of the upstream
flow is greater or smaller than $\omega_{c1}$.  The condition for its existence can be re-expressed as
\be
\womega_{1,\rm X} \simeq 2\gamma_1 \omega \lesssim 2{\gamma_1\over\gamma_2}\omega_{c1}
\ee
in the case of a strong shock ($\gamma_1 \gg \gamma_2$).  The reflected mode is
superluminal when $\womega_{1,\rm X} > \omega_{c1}$ and follows the dispersion relation (\ref{eq:disp_high}).
We will compute the mode amplitude in this regime.

\section{Evaluation of Secondary Modes}\label{s:coupling}

We now solve for the secondary plasma modes in terms of the primary mode that is carried toward the shock
by the upstream flow (Figure \ref{fig:shock}).
We treat the shock as a discontinuity, considering modes with wavelengths greatly exceeding
the downstream particle gyroradius and the shock thickness.
The shock jump can be expressed in terms of the ratio $\gamma_1/\gamma_2 > 1$.

The amplitudes of the secondary modes are obtained by
matching upstream and downstream perturbations at the instantaneous position of the shock.
The analysis that follows makes use of (i) the 
the mode dispersion relations reviewed in Appendix \ref{s:modes_app} and summarized in Section \ref{s:lin_modes}
and (ii) the relations between
mode variables re-evaluated in the frame of the shock, as summarized in Appendix \ref{s:modes_shock}.

\subsection{Shock Jump}\label{s:jump}

We first summarize the flow parameters downstream of a planar, relativistic MHD shock;
for a derivation, see Appendix \ref{s:shock}.
The magnetization is taken to be large and is expressed in terms of the comoving enthalpy
density,\footnote{The expressions that follow are greatly simplified in the case of moderate shock compression by
  assuming the upstream plasma to be relativistically hot in the comoving frame; this guarantees
  that the downstream plasma is also hot.}
$\sigma_1 = B_1^2/4\pi\gamma_1^2w_1\gg 1$.
The downstream flow is also relativistic and strongly magnetized.

The upstream plasma must flow faster than a fast magnetosonic wave in the frame of the shock,
\be
\gamma_1 > \gamma_{\phi,\rm X}(\sigma_1) = \left({3\sigma_1\over 2}\right)^{1/2},
\ee
(see Equation (\ref{eq:gamphiX})).  Then, 
\be\label{eq:gam12}
   {1\over\gamma_2^2} = {1\over\sigma_1} - {1\over 2\gamma_1^2}.  \quad\quad
   \left(\sigma_1 \gg 1\right)
   \ee
In the case of a strong shock,
\be\label{eq:gam2}
\gamma_2 \simeq \sigma_1^{1/2};\quad\quad \sigma_2 \simeq 2\sigma_1.
\ee

Although the ratio $B_2/B_1$ remains close to unity in the shock frame,
there is a compression of the comoving flux density (as measured in the plasma frame):
\be
\wB_2 \simeq \left({\gamma_1\over\gamma_2}\right)\wB_1,
\ee
where $\wB = B/\gamma$.  The quantity $\wB/n$ is conserved exactly in the ideal MHD approximation.

The downstream enthalpy density is 
\be\label{eq:jump}
\begin{split}
w_2 &= w_1 + {B_1^2\over 8\pi}\left({1\over\gamma_2^4}-{1\over\gamma_1^4}\right)\\
&\simeq {1\over 2}\left({\gamma_1\over\gamma_2}\right)^2 w_1. \quad\quad (\gamma_1 \gg \sigma_1^{1/2})
\end{split}
\ee
The second line applies to a strong shock and is obtained by substituting Equation (\ref{eq:gam2}).
The shift in plasma frequency (Equation (\ref{eq:omp})) across the shock is therefore
\be\label{eq:ompratio}
\begin{split}
 {\omega_{p2}\over\omega_{p1}} &= {n_2\over n_1}\left({w_1\over w_2}\right)^{1/2} \\
 &\simeq 2^{1/2}.\quad\quad(\gamma_1 \gg \sigma_1^{1/2})
\end{split}
\ee
         
The effective particle gyrofrequency appearing in the wave dispersion relations is, similarly,
\be
\omega_c = {eB\over {\cal M}c} = \sigma^{1/2}\omega_p
\ee
(see Appendix \ref{s:modes_app}).

The normal flux of transverse momentum (components $i = y,\,z$) is
also conserved across the shock:
\be\label{eq:Txi}
T_{xi,1} = T_{xi,2}.
\ee
The off-diagonal components of the stress-energy tensor are
\be
T_{xi} = uu_iw + {B_xB_i\over 4\pi} + {E_xE_i\over 4\pi}.
\ee
Taking into account that $B_x = 0$ and that all components of ${\bm E}$ are continuous across the shock,
Equations (\ref{eq:jump}) and (\ref{eq:Txi}) reduce to
\be\label{eq:vparallel}
\begin{split}
         {\beta_{i,2}\over\beta_{i,1}} &\simeq \left({\gamma_1\over\gamma_2}\right)^2 {w_1\over w_2} \\
               &\simeq 2.\quad\quad (\gamma_1 \gg \sigma_1^{1/2})
\end{split}
\ee

\subsection{Secondary Mode Amplitudes}\label{s:matching}

We now calculate the amplitudes of the secondary modes.  Our procedure combines two steps.  

(1) For each mode, we express the perturbations to $E$ (the electric field in the plane of the shock),
$\gamma$, $u$, $P$, $n$ in terms of the magnetic perturbation $\delta B$ and phase speed $\beta_\phi$.

(2) Five boundary conditions are applied in the instantaneous frame of the shock.  These are equal in number to the 
conservation equations (\ref{eq:con1})-(\ref{eq:momentum}) and (\ref{eq:vparallel}) but,
when the plasma magnetization is large, some take a much simpler form.

The position of the shock is unperturbed, to linear order, when the incident mode is
incompressible (frozen A-mode).
By contrast, a compressible mode excites a longitudinal motion of the fluid and of the shock; the
flow quantities in the instantaneous shock rest frame are then obtained by an infinitesimal Lorentz
boost of the unperturbed flow by a velocity $\delta\beta_s$.
For example, the perturbations to flow Lorentz factor and four-speed transform as
\be
\delta\gamma \;\rightarrow\; \delta\gamma - u\,\delta\beta_s; \quad\quad
\delta u \;\rightarrow\; \delta u - \gamma\,\delta\beta_s.
\ee
The electric and magnetic field evolve under the same boost as
\be
\begin{split}
\delta\bE &\rightarrow \delta\bE + \delta\beta_s\,\hat x\times\bB;\\
\delta\bB &\rightarrow \delta\bB - \delta\beta_s\,\hat x\times\bE = \delta\bB-\delta\beta_s\,\beta\bB.
\end{split}
\ee

The boundary conditions in the frame of the shock are,
to leading order in inverse powers of $\gamma_1$ and $\gamma_2$,
\be\label{eq:boundary}
\begin{split}
&1. \quad \bE_1 = \bE_2; \\
&2. \quad \gamma_1 n_1  = \gamma_2 n_2; \\
&3. \quad w_1\gamma_1^2\beta_{\perp,1} = w_2\gamma_2^2\beta_{\perp,2};\\
&4. \quad w_2 = w_1 + {B_1^2\over 8\pi}\left({1\over\gamma_2^4}-{1\over\gamma_1^4}\right);\\
&5. \quad {1\over\gamma_2^2} = {1\over\sigma_1} - {1\over 2\gamma_1^2}.
\end{split}
\ee
The fourth and fifth boundary conditions follow from the conservation of energy flux and
normal momentum flux across the shock and are exact in the limit $\sigma_1 \rightarrow \infty$
(see Appendix \ref{s:shock}).  

The boundary conditions (\ref{eq:boundary}) are linearly perturbed, following the
expressions given in Appendix \ref{s:modes_shock}, and expressed in terms of the mode amplitudes $\delta B_i$
on both sides of the shock.  The linear equations so obtained are then solved.

\subsubsection{$A \rightarrow A + O$}

This is the simplest case, because the frozen Alfv\'en mode is incompressible, with
magnetic perturbation $\delta\bB \perp \bB$.  The upstream mode then carries vanishing
perturbations to $\beta$, $\gamma$, $n$, and $w$.  As a result, the perturbations
to the fluxes of particle number, energy and momentum all vanish and the position of
the shock is unperturbed.  The electric perturbations on the upstream and downstream sides are,
from Equations (\ref{eq:dEA}) and (\ref{eq:emrels}),
\be
\begin{split}
\delta\bE_{i,\rm A} &= \beta_i\,\delta B_{i,\rm A}\,\hat y - {\beta_{{\rm A},i}\over\gamma_i}\delta B_{i,\rm A}\,\hat x
\quad (i = 1,2);\\
\delta\bE_{2,\rm O} &= \beta_{\phi,2}\,\delta B_{2,\rm O}\,\hat y.
\end{split}
\ee

Requiring that the normal component of $\delta\bE$ is continuous across the shock gives
\be
\delta B_{2,\rm A} \simeq {\gamma_2\over\gamma_1}\delta B_{1,\rm A} < \delta B_{1,\rm A}.
\ee
Boundary condition 1 applied in the plane of the shock further gives $\delta B_{1,\rm A} \simeq
\delta B_{2,\rm A} + \delta B_{2,\rm O}$.  Hence, a propagating mode that is excited on the downstream
side has an amplitude
\be\label{eq:B2_O}
\delta B_{2,\rm O} \simeq \left(1-{\gamma_2\over\gamma_1}\right)\delta B_{1,\rm A}.
\ee
When the shock is strong, the O-mode carries a large fraction of the fluctuating electromagnetic field
downstream of the shock.  By contrast, the O-mode amplitude vanishes for a very weak shock,
as $\gamma_2\rightarrow \gamma_1$.

\subsubsection{$I \rightarrow I + X$}

The isobaric mode is compressible and excites an oscillation of the shock.  We start by observing
that the reflected extraordinary wave that is emitted by this oscillation -- in the direction
opposite to the relativistic plasma flow -- has a small amplitude $\delta B_{1,\rm X}$ in the frame of the shock.  
We therefore start by setting $\delta B_{1,\rm X} \rightarrow 0$ to determine the amplitudes of the other
modes.  The small but finite value of $\delta B_{1,\rm X}$ is then obtained as a perturbation.

Substituting the transverse electric perturbation $\delta E \simeq -\beta\delta B + \delta\beta_s B$ into 
boundary condition 1 gives
\be\label{eq:pert1}
\begin{split}
\delta B_{1,\rm I} &= \delta B_2 + \delta\beta_s(B_1-B_2) \\
&\simeq \delta B_2 - B_1\left({1\over 2\gamma_2^2} - {1\over 2\gamma_1^2}\right)\delta\beta_s,
\end{split}
\ee
where $\delta B_2 = \delta B_{2,\rm I} + \delta B_{2,\rm X}$.
When perturbing boundary condition 5, we will need $\delta\gamma_I = 0$ for the incompressible $I$ modes
on both sides of the shock
and $\delta\gamma_{2,\rm X}/\gamma_2 \simeq {1\over 2}\delta B_{2,\rm X}/B_2$.
In addition, $\delta\sigma_{1,\rm I}/\sigma_1 = 4\sigma_1\,(\delta B_{1,\rm I}/B_1)$.
Setting $\delta\gamma_1 = -\gamma_1\delta\beta_s$, $\delta\gamma_2 =
\delta\gamma_{2,\rm X} - \gamma_2\delta\beta_s$ and keeping terms to leading order in $\sigma_i$, we get
\be\label{eq:pert2}
\delta\beta_s \simeq -2\sigma_1{\delta B_{1,\rm I}\over B_1}.
\ee
When perturbing boundary condition 4, one has similarly $\delta w_I/w = -4\sigma(\delta B_I/B)$
on both sides of the shock.  Note that $\delta\beta_s$ also involves a factor of $\sigma$,
but $\delta w_{2,\rm X}$ does not (see Equation (\ref{eq:delnX})) and so can be neglected. Then we get to leading order
\be\label{eq:pert3}
   {1\over \gamma_2^2}\delta B_{2,\rm I} = {1\over\gamma_1^2}\delta B_{1,\rm I} + {B_1\over 2}
   \left({1\over\gamma_1^4} - {1\over\gamma_2^4}\right)\delta\beta_s.
\ee

The amplitudes of the downstream modes are expressed in terms of $\delta B_{1,\rm I}$ by combining
equations (\ref{eq:pert1})-(\ref{eq:pert3}),
\be\label{eq:B2_X}
\begin{split}
{\delta B_{2,\rm I}\over\delta B_{1,\rm I}} &= {1 - 3\sigma_1^2/4\gamma_1^4\over 1 - \sigma_1/2\gamma_1^2};\\
{\delta B_{2,\rm X}\over\delta B_{1,\rm I}} &=
-{1 - 3\sigma_1/2\gamma_1^2\over 1 - \sigma_1/2\gamma_1^2}.
\end{split}
\ee
As expected, $\delta B_{2,\rm I} \rightarrow \delta B_{1,\rm I}$ and $\delta B_{2,\rm X} \rightarrow 0$
as the shock becomes very weak, $\gamma_1 \rightarrow \gamma_{\phi,\rm X}(\sigma_1) = (3\sigma_1/2)^{1/2}$.

Now we return to consider the reflected X-mode.  Its amplitude is obtained from the transmitted X-mode wave
by matching the flux of transverse momentum across the shock.  The transverse $e^\pm$ quiver velocity
is obtained from Equation (\ref{eq:betavsE1}), which when evaluated in the frame of the shock is
\be\label{eq:betaperp}
\gamma_1(\delta\beta_\perp)_{1,\rm X}
= i\left({\omega^2-c^2k_1^2\over 4\pi en_1c}\right){\delta E_{1,\rm X}\over\omega}.
\ee
(Here, we have taken into account that the right-hand side of this expression is invariant under Lorentz
boosts along the plasma flow.)  The X-mode has a comoving frequency
\be
\begin{split}
\womega_1 &\simeq 2\gamma_1^2c\wk_{\rm seed} \\
&= 3\sigma_1^{1/2}\left[{\gamma_1\over \gamma_{\phi,\rm X}(\sigma_1)}\right]^2
\left({c\wk_{\rm seed}\over\omega_{p1}}\right)\omega_{c1} \gg \omega_{c1}
\end{split}
\ee
on the upstream side, and a frequency $\womega_2 \simeq \omega/2\gamma_2 \ll \omega_{c2}$ on the downstream side.
The X-mode dispersion relation (\ref{eq:disp}) then gives
\be
\begin{split}
\omega^2-c^2k_{1,\rm X}^2 &= \womega_{1,\rm X}^2-c^2\wk_{1,\rm X}^2 \simeq \omega_{p1}^2; \\
\omega^2-c^2k_{2,\rm X}^2 &\simeq -{2\womega_2^2\over 3\sigma_2}.
\end{split}
\ee
Substituting these expressions into Equation (\ref{eq:betaperp}) and boundary condition 3 gives
\be\label{eq:reflected}
   {\delta B_{1,\rm X}\over \delta B_{2,\rm X}} \simeq - {\delta E_{1,\rm X}\over\delta E_{2,\rm X}} =
   {w_2\over 6\sigma_2 w_1}\left({c\wk_{\rm seed}\over\omega_{p1}}\right)^2.
\ee
Taking into account that the reflected X-mode is present only if $\wk_{\rm seed}$ lies below
the bound (\ref{eq:X1exists}), we have in the case of a strong shock,
\be\label{eq:reflected2}
   {\delta B_{1,\rm X}\over \delta B_{2,\rm X}} \lesssim {1\over 24\sigma_1^2}.
\ee

We conclude that the reflected mode is a small perturbation to the
transmitted modes.  In the fireball model described in Section \ref{s:accel},
a fireball of energy $E \sim 10^{39}$ erg and duration $\sim 10^{-3}$ s has a
magnetization $\sigma_0 \sim 10^4$ in the source zone and
$\sigma_1 \sim 10$ in the interaction zone.  Then the reflected mode
amplitude is minuscule, $\delta B_{1,\rm X}/\delta B_{2,\rm X} \sim 10^{-9} - 10^{-3}$.

A final note: conservation of the $e^\pm$ flow (boundary condition 2) has not been applied here.
That is because the comoving density $n$ does not appear in the other boundary conditions.  Essentially,
boundary condition 2 determines the relative strength of the density and temperature perturbations in
the downstream isobaric mode.  The mode amplitudes are obtained independently of this.

\section{X-mode and O-mode Emission During  Shell Collisions}\label{s:collision}

Intermittency in the electromagnetic outflow from a bursting magnetar can lead to caustic formation
and the creation of shocks in the outflow.  When the magnetization is very large,
the shock strength can be directly related to the amplitude of the variations in Poynting flux.
In this Section, we first describe a simple model of
the forward and reverse shocks produced in an idealized situation involve a collision between
two uniform and strongly-magnetized shells.
This generalizes the treatment of \cite{belob2021b} by allowing for the formation of a reverse shock and
by evaluating the finite compression strength across both shocks.  

Then, we show how the frequencies and powers of the emitted O-mode and
X-mode radiation, as measured in the frame of the magnetar,
are related to those of the seed frozen plasma modes (frozen Alfv\'en wave and
isobaric mode).  These results are applied to a concrete  model of an
accelerating, magnetized fireball in Section \ref{s:accel} and \ref{s:conclusions}, where
the inhomogeneous expansion of the inner,
more luminous shell is also taken into account.

\begin{figure}
\includegraphics[width=\columnwidth]{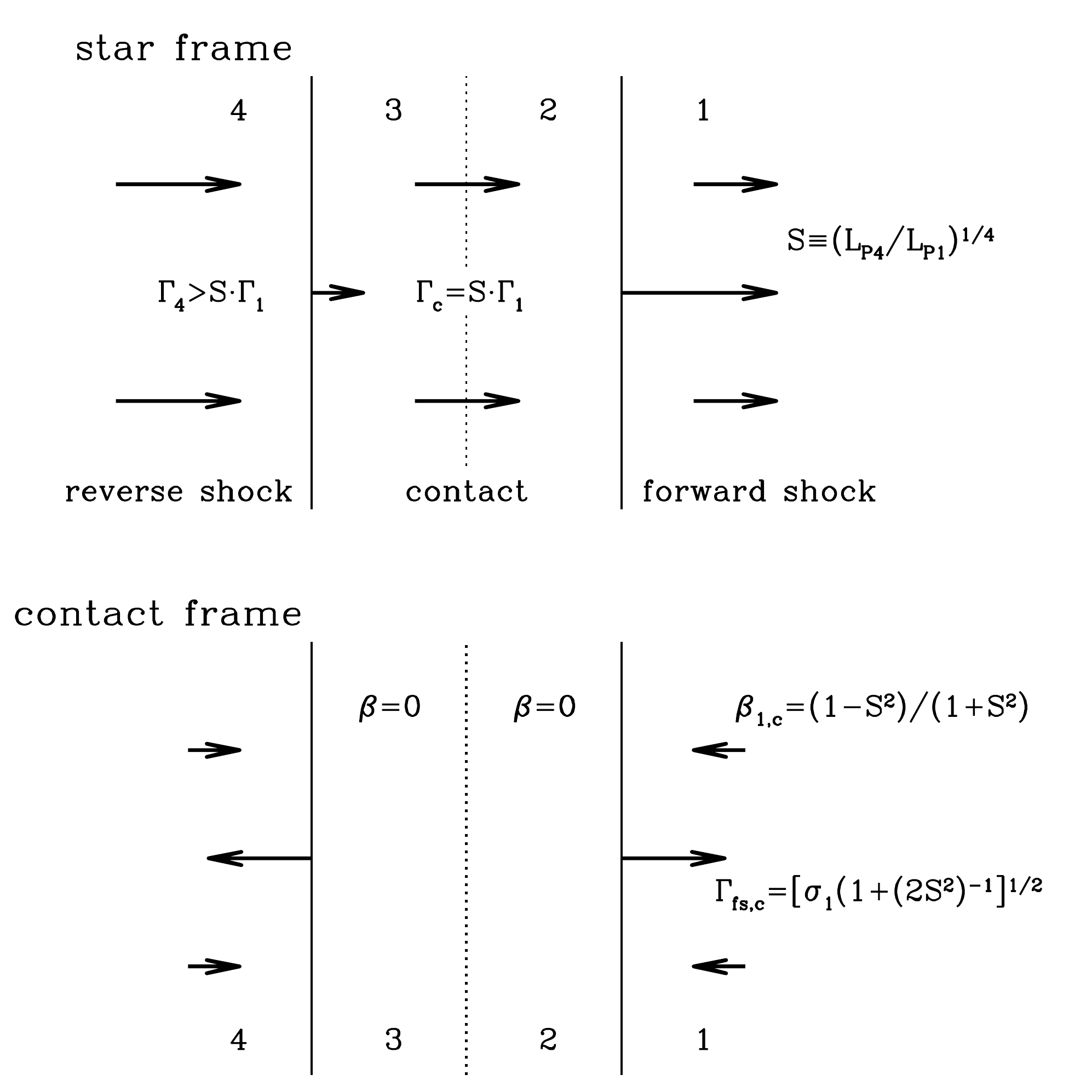}    
    \vskip 0.2in
\caption{Collision of two uniform, relativistically magnetized shells.    The outer, more slowly moving, shell
  is labelled 1 and the inner shell 4;  the motion of both shells is relativistic in the frame of the star,
  $\Gamma_1, \Gamma_4 \gg 1$.   The strength of the forward shock moving into the outer shell (1) is
  determined by the parameter ${\cal S} \equiv (B_4/B_1)^{1/2} = (L_{P4}/L_{P1})^{1/4}$, where the nonradial
  magnetic field $B$ is measured in the frame of the star and $L = B^2 r^2 c$ is the equivalent spherical
  Poynting luminosity.  
  Both the forward and reverse shocks move relativistically in the frame of the contact, so as to satisfy the
  post-shock boundary condition $\beta = 0$ in the shocked layers 2 and 3.
  The shocked material moves with Lorentz factor $\Gamma_c = \Gamma_2 = \Gamma_3 = {\cal S}\Gamma_1$ in the frame of the star,
  which is essentially the center-of-momentum frame defined by the inner and outer electromagnetic flows.
  A strong forward shock requires $L_{P4} \gg L_{P1}$ (but is not required for the linear mode conversion described here);
  a strong reverse shock requires $\Gamma_4 \gg {\cal S}\Gamma_1$.}
\label{fig:shells}
\end{figure}

\subsection{Strength of Shocks Formed by Shell Collisions}\label{s:2shocks}

This late burst of acceleration is ultimately limited by interaction with an exterior medium
that is moving more slowly.  Using a commonly adopted notation (Figure \ref{fig:shells}), we divide the system into four components:
(1) the external medium;  (2) the shocked or compressed external medium;  (3) the shocked or compressed
ejecta shell; and (4) the still freely expanding portion of the ejecta shell.  Layers 2 and 3 are separated
by a contact discontinuity and move with a common Lorentz factor
\be
\Gamma_2 = \Gamma_3 = \Gamma_c.
\ee

We consider the interaction on an intermediate timescale, where the forward and reverse shock have
not yet passed through the impacting layers 1 and 4.  
The forward and reverse shocks move relativistically with respect to the contact, with Lorentz factors
$\Gamma_{\rm fs}$ and $\Gamma_{\rm rs}$ in the frame of the magnetar.
An exact solution is obtained in the
approximation where both slabs are homogeneous: then the Lorentz factor of each shock adjusts
so that the post-shock flow speed matches that of the contact and we have the
ordering
\be\label{eq:Gamineq}
\Gamma_{\rm fs} \gg \Gamma_c;  \quad\quad \Gamma_{\rm rs} \ll \Gamma_c.
\ee
It will be noted that, the forward and reverse shows move relativistically
with respect to the contact, $\Gamma_{{\rm fs},c} = O(\sigma_1^{1/2})$,
$\Gamma_{{\rm rs},c} = O(\sigma_4^{1/2})$, even if $\Gamma_c/\Gamma_1$ and $\Gamma_4/\Gamma_c$
  are not much larger than unity. 

The self-consistent solution for $\Gamma_{\rm rs}$, $\Gamma_c$, and $\Gamma_{\rm fs}$ is obtained
by noting that the electric field is constant across each shock, as measured in the frame of the shock,
\be
(E_2)_{\rm fs} = (E_1)_{\rm fs}; \quad\quad (E_3)_{\rm rs} = (E_4)_{\rm rs}.
\ee
The magnetic pressure is constant across the contact,
\be
(B_2)_c^2 = (B_3)_c^2 \quad\Rightarrow\quad B_2^2 = B_3^2.
\ee
In what follows $B_i$ labels the magnetic field in zone $i$ measured in the frame of the magnetar, and
$E_i = \beta_i B_i$ the corresponding electric field.
Applying the inequalities (\ref{eq:Gamineq}), the magnetic flux densities in the frame of the forward shock are
\be
\begin{split}
(E_1)_{\rm fs} &= \Gamma_{\rm fs}(E_1-\beta_{\rm fs}B_1) \\
           &= \Gamma_{\rm fs}(\beta_1-\beta_{\rm fs})B_1 \simeq -\Gamma_{\rm fs}(1-\beta_1)B_1;\\
  (E_2)_{\rm fs} &\simeq -\Gamma_{\rm fs}(1-\beta_c)B_2.
\end{split}
\ee
We conclude that
\be
B_2 = {1-\beta_1\over 1-\beta_c}B_1 \simeq \left({\Gamma_c\over\Gamma_1}\right)^2B_1.
\ee
The same procedure applied at the reverse shock gives
\be
\begin{split}
(E_4)_{\rm rs} &= \Gamma_{\rm rs}(\beta_4-\beta_{\rm rs})B_4 \simeq {B_4\over 2\Gamma_{\rm rs}};\\
(E_3)_{\rm rs} &\simeq {B_3\over 2\Gamma_{\rm rs}}
\end{split}
\ee
in the regime where the reverse shock is moving away from the star ($\Gamma_{4,\rm rs} \ll \Gamma_4$).
Equating the field pressure on both sides of the contact gives
\be\label{eq:Gamc1}
\Gamma_c = \left({B_4\over B_1}\right)^{1/2}\Gamma_1 = \left({L_{\rm P,4}\over L_{\rm P,1}}\right)^{1/4}\Gamma_1.
\ee
Here $L_{\rm P}$ is the equivalent spherical Poynting luminosity.
One observes that a self-consistent solution is obtained only if the inner shell (4) carries
a higher Poynting luminosity than the outer shell (1).

The strength of the two shocks is also of interest.  The ratio of upstream to downstream Lorentz
factors, measured in the frame of the forward shock, is\footnote{Here $\Gamma_{a,b}$ denotes the
Lorentz factor of zone $a$ as seen in the rest frame of zone $b$.}
\be
   {\Gamma_{1,\rm fs}\over \Gamma_{{\rm fs},c}} = {\Gamma_{\rm fs}/2\Gamma_1\over \Gamma_{\rm fs}/2\Gamma_c} =
     {\Gamma_c\over\Gamma_1}.
\ee
We see from Equation (\ref{eq:Gamc1}) that
the strength parameter for the forward shock is simply related to the luminosity ratio,
\be\label{eq:strength}
   {\cal S} \equiv \left({L_{\rm P,4}\over L_{\rm P,1}}\right)^{1/4}.
\ee
The radio emission process described here does not require a strong forward shock, which in any case
is possible only for a large ratio of luminosities.  
   
We can separately obtain the upstream and downstream Lorentz factors measured in the frame of the shock by
substituting $\gamma_1 \rightarrow \Gamma_{1,\rm fs}$ and $\gamma_2 \rightarrow \Gamma_{\rm fs,c}$
into the jump condition (\ref{eq:gam12}),
\be\label{eq:Gamc_fs}
\Gamma_{1,\rm fs} = \sqrt{\sigma_1\left({1\over 2} + {\cal S}^2\right)};  \quad\quad
\Gamma_{{\rm fs},c} = {\Gamma_{1,\rm fs}\over{\cal S}}.
\ee
As expected, the downstream Lorentz factor $\Gamma_{{\rm fs},c}$
  trends to the strong shock value $\sigma_1^{1/2}$ as $L_{\rm P,1}/L_{\rm P,4} \rightarrow 0$.

Similarly, from the relation
\be
   {\Gamma_{4,\rm rs}\over \Gamma_{{\rm rs},c}} \simeq {\Gamma_4/2\Gamma_{\rm rs}\over\Gamma_c/2\Gamma_{\rm rs}}
   = {\Gamma_4\over\Gamma_c}
\ee
we infer that 
\be\label{eq:Gamc_rs}
\Gamma_{4,\rm rs} = \sqrt{\sigma_4\left({1\over 2} + {\cal S}_{\rm rs}^2\right)};  \quad\quad
\Gamma_{{\rm rs},c} = {\Gamma_{4,\rm rs}\over{\cal S_{\rm rs}}}.
\ee
Substituting Equation (\ref{eq:Gamc1}), the strength parameter for the reverse shock is 
\be\label{eq:Srs}
   {\cal S}_{\rm rs} = {\Gamma_4\over {\cal S}\Gamma_1} = {\Gamma_4\over\Gamma_1}
   \left({L_{\rm P,1}\over L_{\rm P,4}}\right)^{1/4}.
\ee

Our description of the collision as involving two uniform slabs is self-consistent only if the upstream and
downstream flows both move toward the contact (see the bottom panel of Figure \ref{fig:shock}).
A lower bound is implied on the ratio $\Gamma_4/\Gamma_1$.  Requiring that ${\cal S}_{\rm rs} > 1$
gives
\be\label{eq:Gam4min}
\Gamma_4 > \left({L_{\rm P,4}\over L_{\rm P,1}}\right)^{1/4}\Gamma_1.
\ee
When this condition is not satisfied, e.g., when the inner shell is much more luminous than the
outer shell, it is necessarily to consider the inhomogeneous outer profile of the more luminous shell
(Section \ref{s:accel}).  One finds that a relatively thin forward part of the inner shell does
reach the Lorentz factor (\ref{eq:Gam4min}).  This thin contact layer is
continually supplied by material deeper in the shell as it expands and
its mean Lorentz factor continues to grow.

\subsection{O-mode and X-mode Radiation: Star Frame}

The amplitudes of the secondary O-mode and X-mode waves were calculated in Section \ref{s:matching}
in the frame of the shock.   The wave amplitudes and frequencies as observed in the frame of the magnetar
are then obtained by a straightforward Lorentz transformation.

The O-mode and X-mode are modified by this transformation in
significantly different ways.  For a fixed amplitude of the seed zero-frequency mode we find that
\vskip .1in
(1)  The mode with the higher group speed downstream of the forward shock
(the X-mode) emerges significantly weakened compared with the O-mode.
\vskip .1in
(2) Reflection toward the observer enhances the amplitude of the X-mode, but not enough to compensate
the minuscule amplitude of the reflected mode obtained in the frame of the shock.
\vskip .1in
\noindent
Reinforcing these conclusions is the fact that an upstream frozen Alfv\'en mode (which couples
to the O-mode at the shock) typically has a much larger amplitude than the seed isobaric mode
(which couples to the X-mode).

We first consider the radiation emerging from the forward shock and then turn to the case of the reverse shock.

\subsubsection{Forward Shock:  Downstream Modes}

A key point first to be noted is that the downstream O-mode can have a group speed
significantly below the speed of light, as measured in the frame of an emitting shock.
As a result, it propagates toward the observer in the frame of the star.
Refraction by plasma irregularities developing downstream of the shock
  can also boost the frequency and luminosity of the secondary modes, as seen in the frame of the star
  (see Section \ref{s:conclusions}).

The seed magnetic perturbation $\delta B_{\rm seed}$
is carried with the outflow in zone 1, with comoving amplitude $\delta B_{\rm seed}/\Gamma_1$.
The amplitudes in the frame of the magnetar and of the forward shock are related by
\be
(\delta B_{\rm seed})_{\rm fs} = \Gamma_{\rm fs}(1-\beta_{\rm fs}\beta_1)\delta B_{\rm seed} \simeq
\Gamma_{1,\rm fs}{\delta B_{\rm seed}\over\Gamma_1}.
\ee
The downstream mode amplitude $(\delta B_2)_{\rm fs}$ in zone 2 is related to $(\delta B_1)_{\rm fs}$ by
\be
(\delta B_2)_{\rm fs} = F_{\rm shock}({\cal S})\cdot (\delta B_{\rm seed})_{\rm fs}.
\ee
The factor $F_{\rm shock}$ is given by
Equations (\ref{eq:B2_O}) and (\ref{eq:B2_X}) for the O-mode and X-mode, respectively.
Making use of Equation (\ref{eq:Gamc_fs}), this may be written in terms of the variable
${\cal S} = \Gamma_{1,\rm fs}/\Gamma_{\rm fs,c} = (L_{\rm P,4}/L_{\rm P,1})^{1/4}$,
\be
\begin{split}
F_{\rm shock}^{\rm O}({\cal S}) &=  1 - {1\over{\cal S}};\\
F_{\rm shock}^{\rm X}({\cal S}) &= -\left(1 - {1\over{\cal S}^2}\right).
\end{split}
\ee
The downstream electric perturbation is $(\delta E_2)_{\rm fs} = \beta_{g,2}(\delta B_2)_{\rm fs}$, where
$\beta_{g,2} < 0$ and
\be
1-|\beta_{g,2}| = {1-|\wbeta_{g,2}|\over 2\Gamma_{\rm fs,c}^2(1 + |\wbeta_{g,2}|)};
\quad\quad \Gamma_{\rm fs,c} \simeq {\Gamma_{\rm fs}\over 2\Gamma_c}.
\ee
The downstream magnetic perturbation in the frame of the magnetar is then (for either the O- or X-mode)
\be\label{eq:delB2}
\begin{split}
\delta B_2 &= \Gamma_{\rm fs}[(\delta B_2)_{\rm fs} + \beta_{\rm fs}(\delta E_2)_{\rm fs}]\\
&\simeq {\cal S}^2\,F_{\rm shock}({\cal S})\,
\left[{1-|\wbeta_{g,2}|\over 1 + |\wbeta_{g,2}|} + {1\over 4\Gamma_c^2}\right]\cdot\delta B_{\rm seed}.\\
\end{split}
\ee
Here, we have made use of the inequality $\Gamma_{\rm fs} \gg \Gamma_1$ and substituted Equation
(\ref{eq:Gamc1}) for $\Gamma_c$.
The first term in brackets dominates when the wavelength of the O-mode is comparable to the plasma skin depth.
The factor of ${\cal S}^2$ represents reflection from the upstream frame by the frame comoving with the downstream
flow, ${\cal S} \simeq \Gamma_c/\Gamma_1$.   

The frequency of the downstream propagating mode is linearly related to the comoving wavenumber
$\wk_{\rm seed}$ of the seed
zero-frequency mode in zone 1.  The frequency of the seed mode is $\omega_1 = \Gamma_1\beta_1c\wk_{\rm seed}
\simeq \Gamma_1c\wk_{\rm seed}$ in the frame of the magnetar, and
\be
(\omega_1)_{\rm fs} \simeq -{\Gamma_{\rm fs}\over 2\Gamma_1}c\wk_{\rm seed}
\ee
in the frame of the forward shock.  We also have $(\omega_2)_{\rm fs} = (\omega_1)_{\rm fs}$;  hence
the downstream mode has a frequency
\be
\begin{split}
\omega_2 &= \Gamma_{\rm fs}\left[(\omega_2)_{\rm fs} + \beta_{\rm fs}c(k_2)_{\rm fs}\right]\\
&= \Gamma_{\rm fs}\left(1+ {\beta_{\rm fs}\over\beta_{g,2}}\right)(\omega_2)_{\rm fs}.
\end{split}
\ee
in the frame of the magnetar.  Making the same approximations as in the derivation of Equation (\ref{eq:delB2}),
we find
\be\label{eq:om2}
\omega_2 \simeq {\cal S}^2
\left[{1-|\wbeta_{g,2}|\over 1 + |\wbeta_{g,2}|} + {1\over 4\Gamma_c^2}\right]\cdot \Gamma_1c\wk_{\rm seed}.
\ee

We conclude that the transmitted O-mode will typically dominate the transmitted X-mode, for two reasons.
First, the factor $1-|\wbeta_{g,2}| \simeq 1/3\sigma_2$ is small for the transmitted X-mode
(see Equation (\ref{eq:betax})), but is of order
unity in the case of an ordinary wave with $\wk_2 d_2 = O(1)$
(see Equation (\ref{eq:betago})).   Second, a seed isobaric mode is limited
to an amplitude $\delta B_{\rm seed,I}/B \sim 1/\sigma_0$ -- here $\sigma_0$ being the magnetization
at the base of the outflow -- 
whereas a frozen Alfv\'en mode can have a much larger amplitude upstream of the shock.

\subsubsection{Forward Shock:  Reflected X-mode}\label{s:reflect}

The reflected X-mode moves outward with a group speed $\beta_{g,1} \simeq 1$ in the frame of the
forward shock.  Its amplitude is enhanced in the frame of the magnetar compared with that of the shock,
\be
\delta B_{1,\rm X} = \Gamma_{\rm fs}(1+\beta_{\rm fs}\beta_{g,1})(\delta B_{1,\rm X})_{\rm fs} \simeq
          2\Gamma_{\rm fs}(\delta B_{1,\rm X})_{\rm fs}.
\ee
This enhancement is compensated by the small normalization of the reflected X-mode in the
frame of the shock, as given by Equation (\ref{eq:reflected}).  The net result is that the observer sees an
outgoing wave with amplitude
\be
\delta B_{1,\rm X} = F_{\rm shock}^{\rm X}({\cal S})\cdot\Gamma_{1,\rm fs}^2\left({2w_2\over 3w_1\sigma_2}\right)
\left({c\wk_{\rm seed}\over \omega_{p1}}\right)^2\delta B_{\rm seed}.
\ee
The prefactor may be expressed in terms of ${\cal S}$ by substituting
$\Gamma_{1,\rm fs}$ and $\Gamma_{{\rm fs},c}$ for $\gamma_1$ and $\gamma_2$ in Equation (\ref{eq:jump})
and making use of Equation (\ref{eq:Gamc_fs}),
\be\label{eq:dBX1}
\begin{split}
\delta B_{1,\rm X} &= F_{\rm shock}^{\rm X}({\cal S})\,
       {{\cal S}^2(2+{\cal S}^2)^2\over 3(1 + 2{\cal S}^2)}\,
       \left({c\wk_{\rm seed}\over \omega_{p1}}\right)^2\delta B_{\rm seed}\\
       &\simeq {{\cal S}^4\over 6}\left({c\wk_{\rm seed}\over \omega_{p1}}\right)^2\delta B_{\rm seed}.
       \quad ({\cal S} \gg 1)
\end{split}
\ee

The reflected X-mode has a higher frequency, by a factor $\gtrsim \Gamma_{1,\rm fs}^2$, compared with the
downstream X-mode:
\be
\omega_1 = \Gamma_{\rm fs}\left(1 + {\beta_{\rm fs}\over\beta_{g,1}}\right)(\omega_1)_{\rm fs}
\simeq 4\Gamma_{1,\rm fs}^2\cdot\Gamma_1c\wk_{\rm seed},
\ee
giving
\be\label{eq:omX1}
\omega_1 \simeq 2\sigma_1\left(1+2{\cal S}^2\right)\cdot\Gamma_1c\wk_{\rm seed}.
\ee

Although Equations (\ref{eq:dBX1}) and (\ref{eq:omX1}) contain an additional factor of ${\cal S}^2$
as compared with the analogous expressions (\ref{eq:delB2}) and (\ref{eq:om2}) for the transmitted
modes, we must remember that the reflected X-mode is present only when the seed isobaric perturbation has a wavenumber
$\wk_{\rm seed} d_1 \lesssim 1/\Gamma_{1,\rm fs} \sim 1/{\cal S}\sigma_1^{1/2}$ (see Equation \ref{eq:X1exists})).
On balance, $\delta B_{1,\rm X}$ is suppressed compared with $\delta B_{2,\rm O}$ by
a factor $\sim 1/6\sigma_1$ and also by the relative weakness of the seed isobaric mode.

The Poynting flux carried by the reflected X-mode is therefore substantially weaker,
\be
\begin{split}
   {F_{\rm P,O}\over F_{\rm P,X}} &\sim {\delta E_{2,\rm O}\delta B_{2,\rm O}\over
     \delta E_{1,\rm X}\delta B_{1,\rm X}}\\
   &> 36\,\sigma_0^2\sigma_1^2\,
   \left[{1-|\wbeta_{g,2}|\over 1 + |\wbeta_{g,2}|}\right]^2{\delta B_{\rm seed, A}^2\over B^2},
\end{split}
\ee
Here, the amplitude of the seed isobatic mode is limited to $\delta B_{\rm seed,I}/B < 1/\sigma_0$ and
$\wbeta_{g,2}$ is the group speed of the O-mode in the plasma frame.
For example, taking a downstream O-mode wavenumber $\wk_2d_2 = 1$ and group speed
$\wbeta_{g,2} = 1/\sqrt{2}$ along with $\delta B_{\rm seed, A}^2/B^2 \sim (\omega \Delta t)^{-1/2}
= 4\times 10^{-4}\,\nu_9^{-1/2}(\Delta t_{-3})^{-1/2}$ (see the spectrum of seed Alfv\'en modes
given by Equation (\ref{eq:spectrum2})) gives
$F_{\rm P,O}/F_{\rm P,X} > 4\times 10^6(\sigma_0/10^4)^2(\sigma_1/10)^2\,\nu_9^{-1/2}(\Delta t_{-3})^{-1/2}$.
More details about the magnetization of the outflow from a bursting magnetar can be found in
Section \ref{s:accel}.

\subsubsection{Reverse Shock}

It is easier to obtain the observed amplitudes of the secondary modes that are emitted by the reverse shock,
because the required Lorentz boost is now in the direction of the observer.  Taking now
$\delta B_{\rm seed}$ to represent the amplitude of the seed perturbation transported out by the inner shell,
the electric perturbation is  $\delta E_{\rm seed} =  \beta_4 \delta B_{\rm seed}$.
The Lorentz factor of the reverse shock is much smaller
than that of the plasma flows on either side of it;  hence, in the frame of the shock,
\be
(\delta B_{\rm seed})_{\rm rs} = \Gamma_{\rm rs}(1-\beta_{\rm rs}\beta_4)\delta B_{\rm seed} \simeq
       {\delta B_{\rm seed}\over 2\Gamma_{\rm rs}}.
\ee
The secondary modes propagating downstream of the shock have amplitudes
\be
\begin{split}
(\delta B_3)_{\rm rs} &= F_{\rm shock}({\cal S}_{\rm rs})\,(\delta B_{\rm seed})_{\rm rs};\\
(\delta E_3)_{\rm rs} &= \beta_{g,3}\,(\delta B_3)_{\rm rs},
\end{split}  
\ee
where the strength factor ${\cal S}_{\rm rs}$ of the reverse shock is given by Equation (\ref{eq:Srs}).
Boosting back to the frame of the star gives the simple result
\be
\begin{split}
\delta B_3 &= \Gamma_{\rm rs}(1+\beta_{\rm rs}\beta_{g,3})(\delta B_3)_{\rm rs}\\
&\simeq F_{\rm shock}({\cal S}_{\rm rs})\,\delta B_{\rm seed}.
\end{split}
\ee
The observed frequency of the downstream X-mode and O-mode are, similarly,
\be
\omega_3 \simeq \omega_4 = \Gamma_4 c\wk_{\rm seed}.
\ee

\section{Irregular Magnetar Outflow}\label{s:accel}

An irregular outflow from a magnetar provides a context for the shock emission mechanism
described in Sections \ref{s:coupling} and \ref{s:collision}.
We will estimate the Lorentz factor, magnetization, and plasma frequency of the outflow
(which naturally lies in the radio band).
The formation and maintenance of internal shocks is considered, emphasizing the interaction
with a rotationally driven wind.
These results are used in Section \ref{s:conclusions} to compare
the shock-induced O-mode emission with the intrinsic maser-induced emission in the
orthogonal X-mode.

\subsection{Acceleration of a Single Shell}\label{s:accel_single}

Consider an expanding plasma shell whose energy is
dominated by a non-radial electromagnetic field, $(B_\phi, E_\theta)$
in spherical coordinates.  As the shell is released, its ${\bm E}\times{\bm B}$ frame move transrelativistically
outward.  
In the near zone, the stretched poloidal magnetic field connecting to the magnetar
has a significant effect on the shell dynamics, with $B_r \sim B_\phi$ at $r \sim r_0$.
The gyrational frequency of the $e^\pm$ embedded in the shell is several orders of magnitude
larger than the expansion rate, meaning that the bulk velocity is nearly identical to
the ${\bm E}\times{\bm B}$ drift velocity of the embedded charges,
\be
\bbeta \simeq {E_\theta B_\phi\over B^2}\hat r - {E_\theta B_r\over B^2}\hat\theta \simeq {E_\theta\over B_\phi}\hat r
- {B_r\over B_\phi}\hat\theta.
\ee
At large magnetization, this drift quickly becomes relativistic, implying a radial drift Lorentz factor
\be\label{eq:Gamlin}
\Gamma = {1\over\sqrt{1-\beta_r^2}} \sim {B_\phi\over B_r} \sim {r\over r_0}.
\ee
The bulk Lorentz factor increases linearly with radius until saturating at \citep{buckley1977,cerutti2020}
\be
\Gamma_{\rm sat} \sim \sigma_0^{1/3}
\ee
in the case of a steady outflow.
The initial magnetization $\sigma_0$ depends on the plasma state, which will differ significantly
between a quiescent state with voltage set by the flow of corotation charge, and a bursting state
that sustains a broad spectrum of current perturbations.

The shell emitted during an outburst is taken to have a total energy ${\cal E}$, with a fraction
$\varepsilon_\gamma$ carried by photons, and a radial thickness $\Delta r \sim c\Delta t$,
where $\Delta t$ is a characteristic duration (e.g. of X-rays emitted at a modest distance from the magnetar).
The electromagnetic compactness at the release radius $r_0 \sim \Delta r$ is
\be\label{eq:compact}
\ell_{B0} = {\sigma_{\rm T} {\cal E}\over m_ec^2 4\pi r_0^2} = 7.2\times 10^4{{\cal E}_{39}\over (\Delta t_{-3})^2}.
\ee
The effective black body temperature (\ref{eq:teff})
is too low for a significant accumulation of $e^\pm$ in local thermodynamic equilibrium.

In this situation, thermalization in the $e^\pm$ plasma is incomplete \citep{tg2014,belob2021a}.
The scattering depth that develops depends on the relative proportions of energy deposited in thermal and
non-thermal pairs.  For illustration, we assume that heating stops shortly after the release at radius $r_0$.
Then annihilation regulates the scattering optical depth in cold $e^\pm$ to $\tau_{\rm T,0} = O(10)$.
The initial magnetization, as defined by the $e^\pm$ inertia, is
\be\label{eq:sigma0}
\sigma_0 = {\ell_{B0}\over\tau_{\rm T,0}}.
\ee
For a brief interval, the effective magnetization controlling the acceleration of the shell is substantially
smaller, $\sigma \sim \varepsilon_\gamma^{-1}$;  this phase ends as the photons begin to stream
freely with respect to the pairs.

As the ${\bm E}\times{\bm B}$ frame accelerates outward, 
the comoving magnetic field scales as $B'(r) = B/\Gamma \propto (r/r_0)^{-1}\Gamma^{-1} \sim (r/r_0)^{-2}$
and the pair density as $n_\pm(r) \propto (r/r_0)^{-2}\Gamma^{-1} \sim (r/r_0)^{-3}$.
The magnetization decreases intially as
\be\label{eq:siglin}
\sigma(r) = {(B')^2\over 4\pi n_\pm m_ec^2} = {\sigma_0\over\Gamma} \sim \sigma_0\left({r\over r_0}\right)^{-1}.
\ee
   
In the case of strict spherical symmetry \citep{granot2011,lyutikov2010},
the peak Lorentz factor of the shell material can reach a much higher value $\sim 2\sigma_0$
near its forward edge.  The Lorentz factor averaged over the shell
increases as
\be\label{eq:Gamvsr}
\langle\Gamma\rangle(r) \sim \left(\sigma_0 {r\over r_0}\right)^{1/3}.
\ee
This scaling exceeds that given by Equation (\ref{eq:Gamlin}) inside the radius $\sim \sigma_0^{1/2}\,r_0$.
We infer that $\langle\Gamma\rangle$ remains
limited by the stretched poloidal magnetic field until reaching $\langle\Gamma\rangle \sim \sigma_0^{1/2}$.
Beyond this point, the outer part of the shell can continue to accelerate rapidly, but $\langle\Gamma\rangle$
follows the spherical-shell asymptote given by Equation (\ref{eq:Gamvsr}).
The shell thickness remains approximately constant
as long as $\langle\Gamma\rangle \ll \sigma_0$ and the energy flux is dominated by the large-scale electromagnetic field;
hence, the shell-averaged magnetization slowly declines,
\be\label{eq:sigmar}
\langle\sigma\rangle(r) \sim {\sigma_0\over\langle\Gamma\rangle} \sim \sigma_0^{2/3}\left({r\over r_0}\right)^{-1/3}.
\ee

\subsection{Rotationally-Driven Wind}

A burst shell may decelerate through its interaction with a rotationally-driven wind,
or with another shell whose trajectory is influenced by the wind.  In a first approximation,
this wind is quasi-steady, maintaining a nearly uniform magnetization $\sigma_w$ and
asymptotic Lorentz factor $\sim \sigma_w^{1/3}$ \citep{buckley1977,cerutti2020}.
A wind luminosity $L_w$ corresponds to a cross-field voltage $\Phi \sim (L_w/c)^{1/2}$ and magnetization
\be
\sigma_w \sim {e\Phi\over {\cal M}_\pm\gamma_\pm m_ec^2} = 1\times 10^4{L_{w,36}^{1/2}\over
  ({\cal M_\pm})_4 (\gamma_\pm/30)}.
\ee
Here $\gamma_\pm$ is the Lorentz factor of secondary $e^\pm$ produced in a cascade in the open
magnetar circuit and ${\cal M}_\pm$ is the number of such charges per primary corotation charge.

In common with a more luminous burst shell, this wind carries a combed-out radial magnetic field.
The Lorentz factor of its ${\bm E}\times{\bm B}$ frame increases linearly with radius, starting from
a larger launching radius $r_{w0} \sim c/\Omega$ and saturating at $\Gamma_w \sim \sigma_w^{1/3}$ at a radius
$\sim \sigma_w^{1/3}r_{w0}$.  Here $\Omega$ is the angular frequency of rotation of the star.

\subsection{Interaction between Burst Shell and Wind}\label{s:twoshell}

A burst shell expands into the rotationally-driven
wind from a launch radius $r_0 \ll r_{w0}$ (corresponding to a magnetar
spin period much greater than a millisecond).  The shell experiences negligible drag in the outer
corotating magnetosphere, where $B(r) \sim r^{-3}$.  Whether a shell with initial magnetization
$\sigma_{\rm burst,0}$ attains a Lorentz factor $\Gamma_{\rm burst} \gtrsim \sigma_{\rm burst,0}^{1/3}$
inside radius $r_{w0}$ depends on the spin rate of the star;  this happens
if $\sigma_{\rm burst,0} \lesssim (\Omega \Delta t)^{-3} \sim
4\times 10^6\,(P/{\rm s})^3\,(\Delta t_{-3})^{-3}$.  In either case, the shell will move differentially
with respect to the wind in the inner zone where the wind is accelerating outward, as well
as in the saturation zone at $r > \sigma_w^{1/3}r_{w0}$.  

A forward/reverse shock structure forms in the shell as it propagates into the base of the wind.
Here, $\Gamma_{\rm burst}$ can easily satisfy the bound
(\ref{eq:Gam4min}) for a reverse shock to develop.
For example, taking $L_{\rm P,1} = L_w \sim 10^{35}$ erg s$^{-1}$
and a shell luminosity $L_{\rm P,4} = L_{\rm burst} \sim 10^{41}$ erg s$^{-1}$, this bound corresponds to
$\Gamma_4 = \Gamma_{\rm burst} \sim \sigma_{\rm burst,0}^{1/3} > \Gamma_{4,\rm min}$, where
\be\label{eq:Gam4minb}
\begin{split}
\Gamma_{4,\rm min}(r) &\sim \left({L_{\rm burst}\over L_{\rm wind}}\right)^{1/4}\Gamma_w\\
&\sim 30\,{L_{\rm burst,41}^{1/4}\over L_{w,35}^{1/4}}\,{\rm min}
\left[{r\over r_{w0}},\;\sigma_{w0}^{1/3}\right].\quad
(r > r_{w0})\\
\end{split}
\ee

In practice, the shell is inhomogeneous and the expansion Lorentz factor increases strongly
near its forward edge, reaching a maximum value  $\sim 2\sigma_{\rm burst,0}$ when propagating
into a vacuum \citep{granot2011}.
This is easily large enough to sustain a reverse shock in the interior of the shell
as the external wind Lorentz factor increases with $r$.  At first, the increase in $\Gamma_w$
gives the forward part of the burst shell more room for expansion, and the shock structure
drifts toward the front.  After $\Gamma_w$ saturates, 
the average of $\Gamma_{\rm burst}$ over the spreading shell continues to increase and
the shock structure then begins to move backward with respect to the front of the shell.

The radius at which this process is completed can be estimated
by matching Equation (\ref{eq:Gamvsr}) with the critical Lorentz factor (\ref{eq:Gam4minb}),
\be
\begin{split}
r_{\rm rs} &\sim {\sigma_w\over \sigma_{\rm burst,0}}\left({L_{\rm burst}\over L_w}\right)^{3/4}\, r_0 \\
&= 1\times 10^{12}{\sigma_w\over \sigma_{\rm burst,0}}{L_{\rm burst,41}^{3/4}\over L_{w,35}^{3/4}}\Delta t_{-3}
\quad {\rm cm}.
\end{split}
\ee

The Lorentz factor of the contact layer depends weakly on the distribution
of Lorentz factor interior to the reverse shock (see Equation (\ref{eq:Gamc1})).
The magnetic field strength as measured in the frame of the star also does
not vary much across the reverse shock ($B_c \simeq B_4$ in the two-shell model).
After passage of the reverse shock through the shell, the layer of
shocked material (extending to both sides of the contact) has a thickness
\be
\Delta r_{\rm shocked} \simeq 2c\Delta t.
\ee
From this point outward, a rarefaction wave propagates forward, reaching the front of the shell at a radius
\be\label{eq:rdecel}
r_{\rm decel} \simeq {\Delta r_{\rm shocked}\over 1-\beta_c} \simeq 4\Gamma_1^2
\left({L_{\rm P,4}\over L_{\rm P,1}}\right)^{1/2}r_0.
\ee
Here, we have substituted Equation (\ref{eq:Gamc1}) for the Lorentz factor of the shocked shell material
and taken $r_0 \sim c\Delta t$.

The burst shell decelerates outside the radius (\ref{eq:rdecel}).
A concrete example is a rotationally driven wind with magnetization $\sigma_w \sim 10^4$, reaching
a Lorentz factor $\Gamma_1 = \Gamma_w \sim \sigma_w^{1/3} \sim 20$.   Then
\be
r_{\rm decel} \sim 5\times 10^{13}\,\sigma_{w,4}^{2/3}\,{L_{\rm burst,41}^{1/2}\over L_{w,35}^{1/2}}
\Delta t_{\rm burst,-3}\quad {\rm cm}.
\ee
Whether or not deceleration is accompanied by a persistent reverse shock depends on whether
$r_{\rm rs}$ is greater than or smaller than $r_{\rm decel}$, which depends in turn on the relative
magnetization of the wind and the burst shell.

\subsection{Two Closely Spaced Burst Shells}

Consider, finally, the ejection of two relativistic shells, $A$ and $B$,
with comparable luminosities and durations,
separated by an interval $\Delta t_{AB}$.  When $\Delta t_{AB} \sim \Delta t$ (the shell duration),
the two shells are not separated by a significant rotationally-driven wind phase.
Some material in the second shell will move forward over a distance $\sim 2\Gamma_{c,A}^2c\Delta t$
(where $\Gamma_{c,A}$ is the Lorentz factor of the contact in the first shell).
Because $L_{{\rm P},A}\sim L_{{\rm P},B}$, the
equilibrium Lorentz factor of the material in the second shell is comparable
to the Lorentz factor that the first shell attains in its interaction with an outer wind zone,
$\Gamma_{c,B} \sim \Gamma_{c,A}$.  The interaction between the two shells is therefore concentrated
at the same radius $r_{{\rm decel},A}$ that the first shell decelerates.

Alternatively, when $\Delta t_{AB} \gg \Delta t$, the second shell also encounters the wind
and its dynamics is similar to that of the first shell.

\section{Summary and Comparison of Emission Channels}\label{s:conclusions}

We have described a simple linear mechanism producing bright coherent radio emission in
a relativistic outflow from a magnetar -- or other compact star -- with a dynamic magnetic field.
When the magnetization of the outflow is very high, there is a near degeneracy between
subluminal and superluminal expansion.  Small-scale structure that is imprinted into the
magnetic field near the base of the outflow becomes frozen by the expansion and, at
a much greater distance from the star, is directly transformed to superluminal
radiation by a shock.  In the case of magnetars, strong independent evidence for
the presence of small-scale currents during outbursts comes
from the observation of fireball radiation in the X-ray band.

The observed flux of O-mode photons is concentrated around the plasma frequency when
the seed perturbations have a smooth powerlaw spectrum.  This emission occurs 
downstream of the shock, where the ordinary wave has a trans-relativistic group speed when $\wk_2 d_2 = O(1)$.

The two-stage emission mechanism described here has two substantial advantages over direction
emission within a turbulent magnetofluid.  First, turbulent energy is retained and stored below a critical
wavenumber, instead of being continuously channeled into particles.  Second, the direct emission
of fast waves by non-linear coupling to a broad spectrum of Alfv\'en waves becomes inefficient
at a high wavenumber, where the Alfv\'en waves are anisotropic.

As we now demonstrate, this emission channel can compete with or dominate
the intrinsic maser instability of the shock
\citep{plotnikov2019,sironi2021}.  It is weakly sensitive to the strength of the shock and
operates efficiently when the upstream plasma is relativistically hot, in contrast with the
maser \citep{babul2020}.  Strong linear polarization is a natural property of the linear conversion
of a stretched magnetic perturbation to an ordinary wave, but in a direction
orthogonal to that produced by the X-mode maser emission.

A shock propagating through a plasma with extreme magnetization acts as a
very weak reflector, because the plasma flow is relativistic on both sides of the shock.
The amplitude of the reflected X-mode, as observed in the frame of the star,
is found to be minuscule compared to both the downstream ordinary wave and the maser-induced
extraordinary wave.

\subsection{Direct Comparison of Linear O-mode Emission with the X-mode Maser}

We conclude by comparing the brightness of (i) O-mode radiation that emerges downstream of a
shock in response to small current irregularities, and (ii) the intrinsic X-mode maser emission
flowing upstream of the same shock.

Both emission processes include a subdominant component in the orthogonal
polarization.  Isobaric current perturbations seed an X-mode wave downstream of the shock, but their
amplitude is limited in comparison with the frozen Alfv\'en waves that couple to the O-mode.
In the regime of high plasma magnetization, the maser also produces a subdominant flux of O-mode photons
\citep{sironi2021}.  Finally, a negligible X-mode flux is radiated forward of the shock in response to an incoming isobaric
mode (Section \ref{s:reflect}).

The downstream electromagnetic wave is aberrated toward the observer by the relativistic expansion.
For this reason, the detected flux may be significantly modified by refraction downstream of the shock, especially when
$\wk_2d_2 = O(1)$.  Plasma density variations of the required amplitude are expected near a
contact discontinuity forming during a collision  between shells, where the
magnetic field experiences a tearing instability (e.g. \citealt{mahlmann2022}).  To represent
this situation, we consider both the direct emission downstream of the shock, as calculated in
sections \ref{s:coupling} and \ref{s:collision}, and the same radiation field backscattered
elastically away from the star.

The direct O-mode flux  measured in the frame of the star is, following Equation (\ref{eq:delB2}),
\be\label{eq:FPO}
F_{\rm P,O} = {\delta E_{\rm O}\delta B_{\rm O}\over 4\pi}c
\simeq {\cal S}^4\left[{1-|\wbeta_{g,2}|\over 1+|\wbeta_{g,2}|}\right]^2
       {\delta B_{\rm seed}^2\over 4\pi} c.  
\ee
Here, ${\cal S} = (L_{\rm P4}/L_{\rm P1})^{1/4}$ is the
ratio of Lorentz factors across the forward shock that is driven by an outflow of luminosity
$L_{\rm P4}$ into a less luminous outflow of luminosity $L_{\rm P1}$.
Ordinary waves of observed frequency $\omega$ are seeded by Alfv\'enic turbulence of comoving radial
wavenumber $k \simeq \omega/\Gamma c$.  One finds
\be\label{eq:Fseed}
   {F_{\rm P,seed}\over F_{\rm P,1}} = {\delta B_{\rm seed}^2\over B_1^2}
   \sim (\omega \Delta t)^{1-\alpha}
\ee
for a spectrum of seed Alfv\'enic turbulence as given by Equation (\ref{eq:spectrum2}).
The group speed of the ordinary wave is $\wbeta_{g,2} = 1/\sqrt{2}$ when $\wk_2 d_2 = 1$ in the
  downstream rest frame.

Consider next an ordinary wave that is scattered $180^\circ$ into the direction of the outflow,
e.g., near a contact discontinuity forming behind the shock.  We assume that the scattering is
elastic in the frame of the contact, with
\be
\begin{split}
\womega_{2,\rm scatt} &= \womega_2;\quad\quad \wk_{2,\rm scatt} = -\wk_2;\\
\delta\wB_{2,\rm scatt} &= f_{\rm scatt}^{1/2}\delta\wB_2;\quad\quad
\delta\wE_{2,\rm scatt} = -f_{\rm scatt}^{1/2}\delta\wE_2.
\end{split}
\ee
Applying Lorentz boosts to and from the frame of the shock, one finds
\be
\begin{split}
(\omega_{2,\rm scatt})_{\rm fs} &= {1-|\wbeta_{g,2}|\over 1+|\wbeta_{g,2}|} (\omega_2)_{\rm fs};\\
(k_{2,\rm scatt})_{\rm fs} &= -{1-|\wbeta_{g,2}|\over 1+|\wbeta_{g,2}|} (k_2)_{\rm fs},
\end{split}
\ee
and similarly for $(\delta B_{2,\rm scatt})_{\rm fs}$ and $(\delta E_{2,\rm scatt})_{\rm fs}$.
This wavefield moves away from the magnetar both in the frame of the shock and the observer's frame.
Boosting to the latter frame, as in Equation (\ref{eq:delB2}), the amplitude
of the scattered ordinary wave is increased by a factor $\sim 2\Gamma_{\rm fs}$ as compared with the shock frame.
Comparing with the directly emitted wave, one finds
\be
\begin{split}
(\delta B_2)_{\rm scatt} &\simeq f_{\rm scatt}^{1/2}\,(2\Gamma_{\rm fs,2})^2\delta B_2 \\
    &= f_{\rm scatt}^{1/2}\,(2\Gamma_{1,\rm fs})^2\left({1-|\wbeta_{g,2}|\over 1+|\wbeta_{g,2}|}\right) \delta B_{\rm seed}.\\
\end{split}
\ee
Recall that the upstream and downstream Lorentz factors are related by $\Gamma_{1,\rm fs}/\Gamma_{\rm fs,2} = {\cal S}$
(Equation (\ref{eq:Gamc_fs})).  The scattered O-mode energy flux is then
\be\label{eq:FPOscatt}
(F_{\rm P,O})_{\rm scatt} = 16f_{\rm scatt}\,\sigma_1^2\left(1+{1\over 2{\cal S}^2}\right)^2\,F_{\rm P,O}.
\ee
Here, we have evaluated $\Gamma_{\rm fs,2}$ using Equation (\ref{eq:Gamc_fs}).

The upstream temperature of the pairs becomes a significant consideration
  when comparing with the shock maser instability.  The maser output is maximized when the
the upstream pairs are subrelativistic \citep{plotnikov2019,babul2020}.  Then
\be
(F_{\rm P,X}^{\rm maser})_{\rm fs} = {\varepsilon_{\rm maser}\over\sigma_1^2} {(B_1)^2_{\rm fs}\over 4\pi} c
\ee
in the frame of the forward shock; here, the coefficient $\varepsilon_{\rm maser} \simeq 7\times 10^{-4}$.
The O-mode flux is smaller by a factor $\simeq 0.4\sigma_1^{-1}$ when $\sigma_1 \gg 1$ \citep{sironi2021}.
The radiative flux transforms by a factor $\simeq (2\Gamma_{\rm fs})^2
\simeq (4\Gamma_{1,\rm fs}\Gamma_1)^2$ into the frame of the magnetar, giving
\be
F_{\rm P,X}^{\rm maser} \simeq 16\varepsilon_{\rm maser}{\cal S}^4 {B_1^2\over 4\pi} c.
\ee
Here, we have substituted $(B_1)_{\rm fs} = \Gamma_{1,\rm fs}(B_1/\Gamma_1)$ and
$\Gamma_{1,\rm fs} \simeq \sigma_1^{1/2} {\cal S}$.

Comparing the X-mode maser flux with the downstream O-mode flux evaluated in Equation (\ref{eq:FPO})
  using the seed turbulent energy flux of Equation (\ref{eq:Fseed}), we find
\be
   {F_{\rm P,X}^{\rm maser}\over F_{\rm P,O}} \simeq 16\varepsilon_{\rm maser}
   \left[{1-|\wbeta_{g,2}|\over 1+|\wbeta_{g,2}|}\right]^{-2} (\omega\Delta t)^{\alpha-1}.
\ee
The scattered O-mode flux is larger by the factor given in Equation (\ref{eq:FPOscatt}).  The
maser-induced X-mode has a comparable amplitude when the upstream $e^\pm$ are cold,
\be\label{eq:XOratio}
   {F_{\rm P,X}^{\rm maser}\over (F_{\rm P,O})_{\rm scatt}} \;\simeq\; 0.6\,{\nu_9^{1/2}(\Delta t_{-3})^{1/2}\over
   f_{\rm scatt}(\sigma_1/10)^2}.
\ee
This expression corresponds to a strong shock, ${\cal S} \gg 1$, spectral index $\alpha = 3/2$ of the seed frozen modes,
and downstream O-mode wavenumber $\wk_2d_2 \sim 1$.  

The frequency of the downstream O-mode radiation naturally falls in the 100 MHz-GHz range.  For example,
a collision between two shells of luminosity ratio $L_{\rm P,4}/L_{\rm P,1} \sim 10^2$ produces
a forward shock with strength parameter ${\cal S} \sim 3$.  The considerations of Section
\ref{s:twoshell} showed that the collision between may be completed a radius $\sim 10^{14}$ cm.
The Lorentz boosted plasma frequency in the shells is (starting from a Thomson depth $\tau_{\rm T}
\sim 10$ at emission)
\be
 {\Gamma \omega_p\over 2\pi} = 19\,{(\Delta t_{-3})^{1/2}\over r_{14}}\left({\tau_{\rm T,0}\over 10}\right)^{1/2}
   \quad {\rm MHz}.
\ee
The frequency of the direct O-mode wave that is sourced by an Alfv\'en mode of
wavenumber $\wk_{\rm seed} d_1 \sim 1$ is, from Equation (\ref{eq:om2}), larger by a factor
$\sim {\cal S}^2[(1-|\wbeta_{g,2}|)/(1+|\wbeta_{g,2}|)] \sim 1.7(L_{\rm P,4}/10^2\,L_{\rm P,1})^{1/2}$.
The frequency of the scattered O-mode wave is larger than this by an additional factor
$\simeq (2\Gamma_{\rm fs,2})^2 \simeq 4\sigma_1$, giving
\be
   {(\omega_2)_{\rm scatt}\over 2\pi} = 1300\,{(\Delta t_{-3})^{1/2}\over r_{14}}\left({\sigma_1\over 10}\right)\left({{\cal S}\over 3}\right)^2
   \left({\tau_{\rm T,0}\over 10}\right)^{1/2} \quad {\rm MHz}.
\ee

\subsection{Constraint from Induced Compton Scattering}

Induced scattering limits the brightness of the radio emission generated at internal shocks
in an electromagnetic outbursts of the luminosity considered here. (See
\citealt{wr1978} for analogous constraints on the plasma flow from radio pulsars,
and \citealt{lyubarsky2008} for the application to FRBs.)  In the emission zone, the 
energy deposited in radio waves is much smaller than the enthalpy of the frozen $e^\pm$:
$\delta B_{\rm O}^2/B^2 \ll 1/\sigma$.   As a result, the $e^\pm$ are able to absorb the
radio waves with minimal increase in internal energy.

A calculation of the optical depth to induced scattering is complicated by a combination of two effects:
a relativistic energy $\gamma_\pm m_ec^2$ of charges
in the post-shock flow and the cutoff in the spectrum of electromagnetic O-modes below the plasma frequency.
The optical depth at a frequency $\womega$ is suppressed only by a factor $1/\gamma_\pm$ when the spectrum extends
below $\sim \womega/\gamma_\pm^2$ \citep{lyubarsky2008}.  The suppression is stronger (a factor $\sim 1/\gamma_\pm^3$)
in the case of interest where the spectrum is cut off sharply
below the spectral peak (see \citealt{wilson1982} for more general expressions).
Given that the upstream particles have at least a mildly relativistic temperature,
$T_1 \equiv \Theta_1m_ec^2
\gtrsim m_ec^2$, the temperature downstream of a strong shock is $T_2 \simeq w_2/4n_2 \simeq {\cal S}T_1/2$.
(Here, $w_2$ has been evaluated using the jump condition (\ref{eq:jump}) for a strong shock.)
Averaging over a relativistic thermal distribution function gives
\be\label{eq:relfac}
\biggl\langle{1\over\gamma_\pm^3}\biggr\rangle \sim {4\ln({\cal S}\Theta_1/2)\over
  ({\cal S}\Theta_1)^3}.
\ee

To estimate the potential effects of induced scattering, we consider the simpler
  case where the $e^\pm$ are subrelativistic and thermal, with the understanding that the optical depth
  must be rescaled by a factor similar to (\ref{eq:relfac}).  Then
  the time evolution of the photon occupation number $N \equiv kT_b/\hbar\womega$
  can be adequately described by the induced scattering term in the Kompaneets equation \citep{kompaneets1957},
\be
   {\partial N\over\partial t}\biggr|_{\rm ind} = {n_\pm\sigma_T c\over \womega^2}
   {\partial\over\partial\womega}\left({\hbar\over m_ec^2}\womega^4 N^2\right).
\ee   
The optical depth downstream of the shock is, for a photon spectrum $d\ln(\womega^3 N)d/\ln\womega
= -\alpha = -3/2$,
\be\label{eq:tauind0}
\tau_{\rm ind} \equiv \biggl|{\partial\ln N\over\partial \ln t}\biggr|_{\rm ind}
\sim \sigma_T n_2 {r\over\Gamma_2} \cdot \left({T_{\rm b}\over m_ec^2}\right).
\ee
Here, $t \sim r/\Gamma_2 c$ is the time coordinate in the frame of the shocked flow.
   
We now focus on the case where the O-mode radiation generated downstream of the shock
is strongly refracted by plasma inhomogeneities.  Then the mode frequency is
$\womega_2 \sim \omega_{p,2}$ in the comoving frame, and the observed frequency
is $2\pi\nu_{\rm O} \simeq \Gamma_2\womega_{2,\rm O} \sim {\cal S}\Gamma_1\omega_{p,2}$.
The brightness temperature (as measured in the comoving frame)
is related to the observed wave luminosity $L_{\rm O}$ by
\be
4\pi T_{\rm b} \left({\womega_{2,\rm O}\over 2\pi c}\right)^3
\sim {L_{\rm O}\over (4/3)\Gamma_2^24\pi r^2c}.
\ee
Substituting $T_{\rm b}$ into Equation (\ref{eq:tauind0}) gives
\be\label{eq:tauind}
\begin{split}
\tau_{\rm ind} &\sim   {3\sigma_{\rm T}L_{\rm O}\over 64\pi m_ec^3 r}
{m_ec^3/\nu_{\rm O}e^2\over ({\cal S}\Gamma_1)^2}\\
&\sim  {0.5\,L_{\rm O,37}\over ({\cal S}/3)^2\nu_{\rm O,9}r_{14}}
\left({\Gamma_1\over 10^3}\right)^{-2}.
\end{split}
\ee
Induced scattering has only limited effects when the radio wave carries a luminosity
comparable to the brightest event detected from SGR J1935$+$2154.

Further expansion and deceleration of the shocked shell suppresses the plasma frequency relative to the frequency of the
  downstream O-mode radiation.    This enhances the transverse character of the O-mode.
  The wave propagating away from shock does not interact again with the shock near the emission radius;
  the refracted wave only catches up with the shock with some delay, at which point the shell is transparent to the wave.

\subsection{Extension to High Burst Energies}

We have normalized the total energy released by a bursting magnetar in a millisecond period
to the value ${\cal E} \sim 10^{38-39}$ erg appropriate for the radio-emitting pulses
of SGR J1935$+$2154.  Considerably higher luminosities are, of course, associated with magnetar giant
flares \citep{kb2017}.  We conclude by considering how the efficiency of shock-induced radio
emission depends on ${\cal E}$.

Over a wide range of ${\cal E}$, we still expect the scattering depth in $e^\pm$ pairs
at the base of the outflow to be limited by annihilation, $\tau_{\rm T,0} \sim 10$.  
The compactness and magnetization also increase in proportion
to ${\cal E}$, as does the magnetization in the dissipation zone.  This implies an
increased output in O-mode brightness relative to the X-mode maser (Equation (\ref{eq:XOratio})).

On the other hand, Alfv\'en turbulence of a fixed wavenumber $k_\perp \sim k_r$
and coupling strength $k_\perp \delta B_\theta/k_\parallel B_\phi$ carries a current
density $\delta J_\phi \sim (c/4\pi)k_\perp \delta B_\theta \propto {\cal E}^{1/2}$.  After
these modes are frozen by the expansion, the ratio of $\delta J_\phi$ to the maximum current
that can be supplied by the advected charges scales as
\be
   {\delta J_\phi\over en_\pm c} \propto {{\cal E}^{1/2}\over\tau_{\rm T,0}}{r\over \Gamma}.
\ee
If this ratio reaches unity before the plasma experiences a shock, then the advected modes
become charge-starved and damp.  We conclude that the emission radius shrinks and the corresponding plasma
frequency increases with ${\cal E}$, $\omega_p \propto {\cal E}^{1/2}$.   There is therefore
only a limited range of burst energy from which shock-induced ordinary wave radiation in the
100 MHz-GHz range can be released with comparable efficiency, approximately $10^{37} \lesssim {\cal E} \lesssim 10^{41}$ erg.

Induced scattering of the emitted radio waves by the remnant $e^\pm$ fireball pairs sets an
additional constraint on the brightness of the radio waves (Equation (\ref{eq:tauind})).
Setting $\tau_{\rm ind} \sim 1$, the limiting radio luminosity of a pulse of observed duration
$\Delta t_{\rm O} \sim r_{\rm decel}/2\Gamma_c^2c$ (Equation (\ref{eq:rdecel}))
is proportional to $L_{\rm O} \propto r_{\rm decel}\Gamma_c^2 \propto
{\cal S}^4 \Gamma_1^4\Delta t_{\rm O} = (L_{\rm P4}/L_{\rm P1})\Gamma_1^4\Delta t_{\rm O}$.
The Lorentz factors of both colliding shells and of the forward shock increase with the Poynting luminosity
$L_{\rm P}$.  Thus, induced scattering by itself does not rule out radio pulses brighter than the ones
detected from SGR J1935$+$2154, if the underlying flare is also more luminous.

\section*{Acknowledgements}
The author thanks the the Center for Computational Astrophysics for its hospitality when this
work was begun, and especially Luca Comisso, Yuri Levin, Jens Mahlmann, Joonas N{\"a}ttil{\"a},
Alexander Philippov, Yajie Yuan, and Vladimir Zhdankin for stimulating conversations.
The support of the Natural Sciences and Engineering
Research Council of Canada (NSERC) is acknowledged through grant RGPIN-2017-06519.

\section*{Data Availability}
No new data were generated or analysed in support of this research.

\bibliographystyle{mnras}

\appendix

\section{Relativistic MHD Shock}\label{s:shock}

This Appendix reviews the jump of flow variables across a planar shock in a relativistically magnetized
plasma.  We derive
a simple relation between the Lorentz factors on the upstream and downstream sides, which can be used
to quantify the strength of the shock in astrophysical applications.  Although the magnetization in
the upstream flow is assumed to be high, $\sigma = B^2/4\pi \gamma^2 w \gg 1$,
we allow for an arbitrary shock strength.  The formulae obtained
complement the usual analytic approximation positing
a strong shock with arbitrary magnetization \citep{kc1984,zhang2005}.

The upstream and downstream flows are labelled 1 and 2.
The magnetic field ${\bm B}$ is assumed to run parallel to the shock surface;
the comoving enthalpy density $w$ includes the contribution from rest energy.
In this situation, the flow speed $\beta c$ closely approaches the speed of light on both sides of the shock and the
magnetization remains large on on the downstream side.  The mean electric field is directed transverse to the flow,
${\bm E} = -{\bbeta}\times{\bm B}$, with magnitude $|E| = \beta B$.
We work in the frame where the four-velocity ${\bm u} = \gamma\bbeta = \bbeta/\sqrt{1-\beta^2}$ vanishes along
the shock surface.

The relativistic speed of the downstream flow is a consequence of the slow variation with Lorentz factor
of the ratio $\Pi_{\rm P}/S_{\rm P}$.  Here
$S_{\rm P} = |E|Bc/4\pi = \beta B^2c/4\pi$ is the Poynting flux in the direction of the flow
and $\Pi_{\rm P} = (B^2 + E^2)/8\pi = (1+\beta^2)B^2/8\pi$
the electromagnetic momentum flux.  The jump condition $E_1 = E_2$
(as derived from Amp\`ere's law) implies
\be\label{eq:con1}
B_2 = {\beta_1\over \beta_2}\,B_1.
\ee
Continuity of the particle flux $nu$ (here $n$ is comoving particle density) also gives
\be\label{eq:con2}
n_2 = {\gamma_1\beta_1\over \gamma_2\beta_2}\,n_1.
\ee
Balancing the total energy flux and momentum flux across the shock further implies
\be\label{eq:energy}
\gamma_1u_1 w_1 + \beta_1{B_1^2\over 4\pi} = \gamma_2u_2 w_2 + \beta_2{B_2^2\over 4\pi}
\ee
and
\be\label{eq:momentum}
u_1^2 w_1 + P_1 + {B_1^2\over 8\pi} = u_2^2 w_2 + P_2 + {B_2^2\over 8\pi}.
\ee
Here, $P$ is the (comoving) thermal pressure.  

Working in the regime $\sigma_1 \gg 1$ and $\gamma_1 \gg 1$, we take the 
difference of Equations (\ref{eq:energy}) and (\ref{eq:momentum}).  The kinetic terms on the
left-hand side nearly cancel:  the term $P_1 \propto 1/\sigma_1$ and the other difference terms
scale as $1/\gamma_1^2$.  Hence a strong shock requires $\gamma_1^2 \gg \sigma_1$.
The derivation is simplified by the assumption of a relativistically hot upstream flow,
corresponding to $w_1 = 4P_1$; this guarantees that the downstream flow is also hot, even
if the shock is weak.

Expanding in powers of $1/\gamma_1^2$ and $1/\gamma_2^2$, one finds to leading order
\be\label{eq:w2}
\begin{split}
w_2 &\simeq w_1 + {B_1^2\over 2\pi}\left[\left(1-{\beta_1\over\beta_2}\right)^2
  - 2(1-\beta_1)\left(1-{\beta_1\over\beta_2}\right)\right]\\
&\simeq w_1 + {B_1^2\over 8\pi}\left({1\over\gamma_2^4}-{1\over\gamma_1^4}\right).
\end{split}
\ee
Substituting this expression for $w_2$ into either Equation (\ref{eq:energy}) or (\ref{eq:momentum})
gives, to the same order,
\be
\left(\gamma_1^2-\gamma_2^2\right)\left[\left({\gamma_1\over\gamma_2}\right)^2 + {1\over 2} -
    {\gamma_1^2\over\sigma_1}\right] = 0.
\ee
One solution to this equation corresponds to a continuous flow, and the other to a shock jump
\be
   {1\over\gamma_2^2} = {1\over\sigma_1} - {1\over 2\gamma_1^2}  \quad\quad
   \left(\sigma_1 \gg 1;\;\; \gamma_1 > \sqrt{3\sigma_1\over 2}\right).
\ee
The solution for $w_2$ is found by substituting this expression into Equation (\ref{eq:w2}).

The downstream Lorentz factor is seen to vary only over a narrow range, $\sigma_1^{1/2} < \gamma_2 <
\gamma_{\phi,\rm X}(\sigma_1) = (3\sigma_1/2)^{1/2}$ (see Equation (\ref{eq:gamphiX})).
The shock is weak, $\gamma_1/\gamma_2 \gtrsim 1$, when the upstream flow moves only slightly
faster than the fast magnetosonic mode.  The strong-shock asymptote is
\be
\gamma_2 \simeq \sigma_1^{1/2};\quad  w_2 \simeq {B_1^2\over 8\pi \sigma_1^2}.
\ee
so that
\be\label{eq:wjump}
w_2 \simeq {1\over 2}\left({\gamma_1\over\gamma_2}\right)^2 w_1 \gg w_1;
\quad\quad {w_2\over n_2} \simeq \left({\gamma_1\over 2\gamma_2}\right){w_1\over n_1}.
\ee
The downstream magnetization is even larger than on the upstream side:  
\be
\sigma_2 = {B_2^2\over 4\pi \gamma_2^2 w_2} \simeq 2\gamma_2^2 \simeq 2\sigma_1.
\ee
Although the comoving temperature
rises behind the shock, so does the comoving magnetic flux density $B/\gamma$.

\section{Electromagnetic Modes:  Transverse Propagation}\label{s:modes_app}

We now review the dispersion relations of the ordinary and extraordinary electromagnetic modes
(O-mode and X-mode) in a strongly magnetized and relativistic $e^\pm$ plasma.  
Along the way, we obtain relations between the wave variables that are needed in the
study of shock perturbations in Section \ref{s:coupling}.  
We consider only a charge-balanced $e^\pm$ gas and assume that the background plasma is at rest.
The dispersion relations in a plasma in uniform motion are easily obtained by applying a Lorentz boost
to the results obtained below (see Section \ref{s:modes}).

The X-mode can be given a hydromagnetic description at frequencies below the particle gyrofrequency,
being identified with the compressive fast magnetosonic mode.  The O-mode propagates only above
the plasma frequency and is effectively incompressible because the magnetic perturbation
$\delta\bB \perp \bB$.  The O-mode may be excited
at a shock by a frozen, transverse perturbation (essentially, a low-frequency Alfv\'en mode)
that is advected with the upstream plasma.  The compressible X-mode is excited
when a zero-frequency isobaric mode (entropy mode) collides with the shock.

In our study of shock perturbations, the
mean magnetic field runs parallel to the shock surface and the mode wavevector ${\bm k}$
points normal to this surface and to ${\bm B}$ (see Section \ref{s:modes}).   We choose
coordinates ${\bm k} = k\,\hat x$ and ${\bm B} = B\,\hat y$.  The perturbation is Fourier decomposed as
\be\label{eq:fourier}
\begin{split}
\delta{\bm X} &= \delta{\bm X}_0 e^{i(kx-\omega t)};\\
\delta{\bm X} &\equiv \{\delta{\bm B},\, \delta{\bm E},\, \delta\bbeta,\, \delta P,\, \delta n\}.
\end{split}
\ee
The electric and magnetic perturbations are related by Faraday's law,
\be\label{eq:faraday}
\delta{\bm B} = {c\over\omega}{\bm k}\times \delta{\bm E}  = {1\over\beta_\phi}\hat x\times\delta{\bm E}.
\ee
The phase speed $\beta_\phi = \omega/ck$ can have either sign.

Positive and negative charges are oppositely accelerated along $\delta{\bm E}$, gaining
a quiver velocity $\pm\bbeta_E$, but experience a Lorentz force $(\pm e)(\pm\delta\bbeta_E)\times{\bm B}$
of the same sign:
\be\label{eq:beta}
\delta\bbeta_\pm = \pm \delta\bbeta_E + \delta\bbeta_{E\times B}.
\ee
The current is
\be
\delta{\bm J} = en_+\delta\bbeta_+ - en_-\delta\bbeta_- = en\delta\bbeta_E.
\ee
where $n = n_+ + n_-$ is the total space density of positrons and electrons.  The corresponding Maxwell equation
is
\be\label{eq:maxwell}
-i\omega\delta{\bm E} = -4\pi \delta{\bm J} + ikc(\hat x\times\delta{\bm B}).
\ee

We treat the positive and negative particles as fluids with
the same space density, pressure, enthalpy density $w$, and effective mass
\be
   {\cal M} = {w_+\over n_+ c^2} = {w_-\over n_- c^2} = {w\over nc^2}.
\ee
The inertial mass density is $n{\cal M} = w/c^2$.  The linearized momentum equations for positrons and electrons,
\be\label{eq:mom}
   {w_\pm\over c}{\partial(\delta\bbeta_\pm)\over\partial t} = \pm en_\pm\left(\delta{\bm E} + \delta\bbeta_\pm\times{\bm B}\right) -
   \bnabla \delta P_\pm.
   \ee
Here, $e$ is the magnitude of the electron charge.
The pressure gradient is evaluated in the adiabatic approximation,
$\partial P/\partial x = (\partial P/\partial n)_S \cdot \partial n/\partial x$.  Combining this
with the linearized continuity equation,\footnote{Creation and annihilation of $e^\pm$ pairs
  is negligible over the wave period.}
\be\label{eq:contin}
   {\partial(\delta n)\over\partial t} = -nc{\partial(\delta\beta_x)\over\partial x},
\ee
the sound speed is given by
\be
c_s^2 = {1\over{\cal M}}\left({\partial P\over \partial n}\right)_S.
\ee
The case of a relativistic $e^\pm$ gas corresponds to
$w = 4P$, $c_s = c/\sqrt{3}$ and (when the distribution function is thermal) ${\cal M} = 4kT/c^2$.

We now examine separately the two transverse electromagnetic modes propagating perpendicular
to ${\bm B}$.

\subsection{X-mode (Fast Magnetosonic Mode)}

The X-mode carries an electric perturbation directed along ${\bm k}\times{\bm B}$;  hence,
\be\label{eq:xmode}
   {\bm k} = k\,\hat x;  \quad
   \delta {\bm E} = \delta E\,\hat z; \quad \delta {\bm B} = \delta B\,\hat y;  \quad\quad ({\rm X})
\ee
\be
\delta\bbeta_{E\times B}  = \delta\beta_{E\times B}\,\hat x.   \quad\quad ({\rm X})
\ee
The momentum equations (\ref{eq:mom}), written in terms of the variables (\ref{eq:beta}), are
\be\label{eq:mom1}
         {\partial(\delta\beta_E)\over\partial t} = {e\over{\cal M}c}(\delta E + \delta\beta_{E\times B}\, B);
\ee
\be\label{eq:mom2}
{\partial(\delta\beta_{E\times B})\over\partial t} = -{eB\over{\cal M}c}\delta\beta_E -
{1\over {\cal M}nc}{\partial(\delta P)\over\partial x}.
\ee
Here, $\delta P = \delta P_+ + \delta P_-$.

We next substitute Equations (\ref{eq:fourier}) and (\ref{eq:xmode})
into Equations (\ref{eq:faraday}), (\ref{eq:maxwell}), (\ref{eq:contin}) and (\ref{eq:mom2})
to get following relations between flow variables 
\be\label{eq:betavsE1}
\delta\beta_E =  i \left({\omega^2-c^2k^2\over 4\pi enc}\right){\delta E\over \omega}
= i \sigma^{1/2}\left({\omega^2-c^2k^2\over \omega \omega_p}\right){\delta E\over B};
\ee
\be\label{eq:betavsE2}
\delta\beta_{E\times B} = \sigma\left({\omega^2-c^2k^2\over \omega^2 - c_s^2k^2}\right){\delta E\over B}.
\ee
Here, $\sigma = B^2/4\pi n{\cal M}c^2$; the effective plasma frequency and cyclotron frequency are
\be
\omega_p = \left({4\pi ne^2\over {\cal M}}\right)^{1/2};  \quad\quad  \omega_c = {eB\over{\cal M}c} = \sigma^{1/2}\omega_p.
\ee

The dispersion relation is obtained by substituting Equations (\ref{eq:betavsE1}) and (\ref{eq:betavsE2})
into Equation (\ref{eq:mom1}),
\be\label{eq:disp}
\omega^2-c^2k^2 = 
-{\omega_p^2(\omega^2-c_s^2k^2)\over \omega_c^2 - \omega^2 + c_s^2k^2}.
\ee
The mode is subluminal for $\omega \lesssim \omega_c$; at low frequencies, the dispersion relation
approaches the familiar form given by Equation (\ref{eq:betax}).
A separate superluminal branch asymptotes to the unmagnetized dispersion relation
\be\label{eq:disp_high}
\omega^2 = c^2k^2 + \omega_p^2.\quad\quad (\omega \gg \omega_c)
\ee
at high frequencies.

The $e^\pm$ are tied to the magnetic field at low frequencies.  Then, the mode quiver speed $\beta_{E\times B}$
coincides with the quasi-static ${\bm E}\times{\bm B}$ drift speed,
\be\label{eq:betavsE}
\delta\beta_{E\times B} \simeq \hat x\cdot{\delta{\bm E}\times{\bm B}\over B^2} = -{\delta E\over B},
\ee
as may be seen by substituting Equation (\ref{eq:disp}) in Equation (\ref{eq:betavsE2}).  Hence
\be\label{eq:betavsB}
\delta\beta_{E\times B} \simeq \beta_\phi {\delta B\over B}.
\ee
The transverse quiver is suppressed by a factor $\sim \omega/\omega_c$,
\be
\delta\beta_E \simeq i \left(1-{c_s^2\over c^2}\right)\left({\omega\over\omega_c}\right){\delta E\over B}.
\ee

\subsection{Low-frequency O-mode}

The polarization of the O-mode is orthogonal to that of the fast magnetosonic mode;  hence
\be\label{eq:omode}
   {\bm k} = k\,\hat x;  \quad
   \delta {\bm E} = \delta E\,\hat y; \quad \delta {\bm B} = \delta B\,\hat z.  \quad\quad ({\rm O})
   \ee
   The analysis is now much simpler, because the low-frequency oscillation decouples from the Lorentz force.
   The $y$-component of the momentum equation reduces to
\be\label{eq:mom1b}
         {\partial(\delta\beta_E)\over\partial t} = {e\over{\cal M}c}\delta E;  \quad\quad
         \delta\beta_E = i\left({\omega_c\over\omega}\right){\delta E\over B}.
\ee
Combining this with the Faraday and Maxwell equations
(\ref{eq:faraday}) and (\ref{eq:maxwell}) gives
\be
\omega^2 = c^2k^2 + \omega_p^2.
\ee
The mode phase speed is superluminal,
\be
\beta_\phi = \left(1+{\omega_p^2\over c^2k^2}\right)^{1/2}.
\ee
Particle gyromotion is excited in a distinct, high-frequency branch of the dispersion relation,
$\omega \sim \omega_c$.   

\subsection{Very Low-Frequency Shear Alfv\'en Mode}

The final mode considered in this Appendix has a much lower frequency than the other two.
Both the magnetic and electric fluctuations are now transverse to ${\bm B}$,
\be\label{eq:amode}
   {\bm k} = k\left(\hat x + \varepsilon\hat y\right);  \quad
   \delta {\bm E} = \delta E\,\hat x; \quad \delta {\bm B} = \delta B\,\hat z.  \quad\quad ({\rm A})
   \ee
   The current supporting the magnetic fluctuation runs parallel to ${\bm B}$, as with the O-mode,
   but now the electric perturbation only induces slow ${\bm E}\times{\bm B}$ drift,
\be\label{eq:betaA}
\begin{split}
\delta\bbeta_\pm &= \pm{k\delta B\over 4\pi en}\hat B + \delta\bbeta_{E\times B}\\
   &= \pm\sigma^{1/2}\left({kc\over\omega_p}\right){\delta B\over B}\hat y + {\delta E\over B}\hat z.
\end{split}
\ee
This mode is the very low-frequency limit of a shear Alfv\'en wave propagating along ${\bm B}$
with phase speed $\beta_{\rm A} = (1+1/\sigma)^{-1/2}$, parallel wavevector $k_\parallel = \varepsilon k \ll k$,
and a frequency $\omega = \beta_{\rm A}ck_\parallel$ which vanishes
as $\varepsilon \rightarrow 0$.   The electric fluctuation is obtained from Faraday's law,
\be
\delta{\bm E} =-\beta_{\rm A}\,\delta B\,\hat x.
\ee
(One must include the small $y$-component of ${\bm k}$ in Equation (\ref{eq:faraday})
to obtain the correct answer.)

\section{Mode Variables in the Shock Frame}\label{s:modes_shock}

In this Section, we translate into the (unperturbed) shock frame the relations between mode variables 
described in Section \ref{s:modes} and Appendix \ref{s:modes_app}.  The subscripts
A, I, O and X label the frozen Alfv\'en mode, the isobaric mode, the ordinary
electromagnetic mode, and the extraordinary mode (fast magnetosonic mode).  As in the main text,
a quantity labeled with a tilde is evaluated in the plasma rest frame.

(1) Zero-frequency modes (A and I), $\womega_{\rm A,I} = 0$.  In the comoving
frame, these modes have vanishing phase speed $\wbeta_\phi$; hence their phase is tied to the mean flow
in the shock frame:
\be
\beta_\phi = {\omega\over ck} = {\wbeta_\phi + \beta\over 1 + \beta\wbeta_\phi} = \beta. \quad\quad ({\rm A,I})
\ee
The electric perturbation transverse to $\wbk$ also vanishes in the plasma frame.
Lorentz transforming to the frame of the shock, the transverse magnetic perturbation is therefore
$\delta\bB_{\rm A,I} = \gamma\delta\wbB_{\rm A,I}$ and
\be
   {\delta B_{\rm A,I}\over B} = {\delta\wB_{\rm A,I}\over\wB}.
\ee
The electric perturbation in the shock frame is
\be\label{eq:dEA}
\begin{split}
\delta\bE_{\rm A} &= -\beta\,\hat x\times\delta\bB_{\rm A} - {\beta_{\rm A}\over\gamma}\delta B_{\rm A}\,\hat x;\\
\delta\bE_{\rm I} &= -\beta\,\hat x\times\delta\bB_{\rm I}.
\end{split}
\ee
The frozen A-mode carries a longitudinal electric field ($\parallel\,\wbk$) that is invariant between frames and
is preserved across the shock;  it contributes weakly to the comoving electromagnetic field
downstream of a strong shock.

The isobaric mode has a finite pressure perturbation
\be
\begin{split}
   {\delta P_{\rm I}\over P} &= -{\wB\,\delta\wB_{\rm I}\over 4\pi P}\\
                       &= -\sigma{w\over P}{\delta\wB_{\rm I}\over\wB} = -\sigma{w\over P}{\delta B_{\rm I}\over B}.
\end{split}
\ee
The corresponding comoving density perturbation is
\be
{\delta n_{\rm I}\over n} = {\delta P_{\rm I}\over P} - {\delta T_{\rm I}\over T}.
\ee
The isobaric mode also has a vanishing velocity perturbation in the comoving frame, and therefore
in the shock frame:
\be
\delta\bbeta_{\rm I} = \delta\widetilde{\bbeta}_{\rm I} = 0.
\ee

The frozen Alfv\'en mode is incompressible and so 
\be
\delta P_{\rm A} = \delta n_{\rm A} = 0.
\ee
Upstream of the shock, the differential
$e^\pm$ drift along $\wbB$ supports a current perturbation that satisfies the steady Maxwell equation.
Translating Equation (\ref{eq:betaA}) to the shock frame gives
\be
\delta\beta_{y,\rm A} = {1\over\gamma^3}{k_{\rm A}\delta B_{\rm A}\over 4\pi en}.
\ee
This mode has vanishing $\delta\widetilde\bbeta$ along $\wbk$ (see Equation (\ref{eq:betaA})), consistent
with its being tied to the mean flow. 

2. Finite-frequency modes (O and X).  Now the mode supports an electric perturbation in the plasma
rest frame,
\be\label{eq:emrels}
\delta\wbE = -\wbeta_\phi\,\hat x\times\delta\wbB.
\ee
The mode phase speed $\wbeta_\phi = \womega/c\wk$ 
is given by Equations (\ref{eq:betao}) and (\ref{eq:betax}) in the comoving frame;  translating 
to the shock frame gives
\be\label{eq:bphish}
1-\beta_\phi \simeq {1-\wbeta_\phi\over 2\gamma^2(1+\wbeta_\phi)}.\quad\quad (\gamma \gg 1)
\ee
The electric and magnetic perturbations are both transverse to $\bk$, and
are related by Equation (\ref{eq:emo}) in both the plasma frame and the shock frame.
In the (unperturbed) shock frame, the electromagnetic perturbation is is transformed to
\be
\delta\bB = \gamma(1+\beta\wbeta_\phi)\delta\wbB;  \quad
\delta\bE = -\beta_\phi\,\hat x\times\delta\bB.\quad\quad ({\rm O,X})
\ee
Hence,
\be
   {\delta B\over B} = (1+\beta\wbeta_\phi){\delta\wB\over\wB} \simeq 2{\delta\wB\over\wB}.\quad\quad({\rm O,X})
\ee   

The O-mode is incompressible; hence
\be
\delta n_{\rm O} = \delta P_{\rm O} = 0.
\ee
The velocity perturbation is along $\bB$;  Lorentz transforming to the shock frame gives an expression
identical in form to Equation (\ref{eq:mom1b}),
\be
\delta\beta_{y,\rm O} = i\left({\omega_c\over\omega}\right){\delta E\over B} =
i\beta_\phi\left({\omega_c\over\omega}\right){\delta B\over B}.
\ee

The X-mode is compressible but adiabatic.
The longitudinal ${\bm E}\times{\bm B}$ velocity
perturbation shifts the Lorentz factor of the upstream plasma flow,
\be
\begin{split}
\gamma &\;\rightarrow\; \gamma + \delta\gamma_{\rm X} = \gamma + \delta\wbeta_{\rm X} u;\\
u &\;\rightarrow\; u + \delta u_{\rm X} = u + \delta\wbeta_{\rm X} \gamma.
\end{split}
\ee
This kinetic perturbation is not present in the I, A, or O modes -- in these cases, $\delta\beta = 0$.
The equation of continuity (Equation (\ref{eq:contin})) implies
\be
   {\delta n_{\rm X}\over n} = {\delta\wbeta_{\rm X}\over\wbeta_{\phi,\rm X}},
\ee
where from Equations (\ref{eq:betavsE}) and (\ref{eq:betavsB}),
\be
\delta\wbeta_{\rm X} = -{\delta\wE_{\rm X}\over\wB} = \wbeta_{\phi,\rm X}{\delta\wB_{\rm X}\over\wB}.
\ee
Hence,
\be\label{eq:delnX}
   {\delta n_{\rm X}\over n} = {1\over\Gamma}{\delta P_{\rm X}\over P}
     = {1\over 1+\beta\wbeta_\phi}{\delta B\over B},
\ee
where $\Gamma$ is the ratio of specific heats.  Because the X-mode is essentially a MHD wave at
low frequencies ($\womega \ll \omega_c$),  one may also write $\bE = -\beta\hat x\times\bB$ and perturb to get
\be
\delta E_{\rm X} = -\beta \delta B_{\rm X} - {\delta\wbeta_{\rm X}\over\gamma^2}B.
\ee

\bsp	
\label{lastpage}
\end{document}